\documentclass[aps,prb,twocolumn,groupedaddress]{revtex4}
\usepackage{graphicx}
\usepackage{dcolumn}
\usepackage{bm}
\usepackage{amsmath,amsfonts,amssymb,amsthm,verbatim}
\usepackage[ansinew]{inputenc}
\usepackage[usenames]{pstricks}
\usepackage{color}
\usepackage{epsfig}
\usepackage{MnSymbol}
\usepackage{bbold}
\usepackage{dsfont}
\usepackage{hyperref}
\hypersetup{
     colorlinks=true,
     linkcolor=black,
     filecolor=black,
     citecolor = blue,      
     urlcolor=blue,
     }

\newcommand{\be}{\begin{equation}}
\newcommand{\ee}{\end{equation}}
\newcommand{\bea}{\begin{eqnarray}}
\newcommand{\eea}{\end{eqnarray}}

\newcommand{\bi}{\begin{itemize}}
\newcommand{\ei}{\end{itemize}}
\newcommand{\bc}{\begin{center}}
\newcommand{\ec}{\end{center}}

\begin{document}

\title{Topology of critical chiral phases: Multiband insulators and superconductors}

\author{Oleksandr Balabanov$^{1, 2}$}
\author{Daniel Erkensten$^{3}$}
\author{Henrik Johannesson$^{2}$}
\affiliation{$^{1}$Department of Physics, Stockholm University, SE 106 91 Stockholm, Sweden}
\affiliation{$^{2}$Department of Physics, University of Gothenburg, SE 412 96 Gothenburg, Sweden}
\affiliation{$^{3}$Department of Physics, Chalmers University of Technology, SE 412 96 Gothenburg, Sweden}

\begin{abstract}
Recent works have proved the existence of symmetry-protected edge states in certain one-dimensional topological band insulators and superconductors at the gap-closing points which define quantum phase transitions between two topologically nontrivial phases. We show how this picture generalizes to multiband critical models belonging to any of the chiral symmetry classes AIII, BDI, or CII of noninteracting fermions in one dimension.
\end{abstract}


\maketitle

\section{Introduction}

The presence of topological edge states, decoupled from the bulk, is a key characteristic of symmetry-protected topological phases of quantum many-body systems \cite{Chiu2016}. In one dimension (1D), these states are exponentially localized at the physical boundaries of the system, making their energy vanish identically in the thermodynamic limit (and sometimes also at certain fine-tuned points in the phase diagram of a finite system). With no interactions present, the possible 1D fermionic topological phases are those of the topological band insulators and mean-field superconductors \cite{Hasan,Qi}, classified by the "ten-fold way" \cite{Kitaev2009,Schnyder2008,Ryu2010}. For this class of models the very existence of edge states is a consequence of the topologically nontrivial phase structure of the single-particle bulk states ("bulk-boundary correspondence" \cite{Teo2010}), with their robustness against local perturbations (or uncorrelated disorder) being ensured by the symmetries enforced on the perturbations. Well-known examples include the fractionalized soliton mode of the Su-Schrieffer-Heeger model \cite{SSH1979} and the Majorana zero-energy mode of the Kitaev chain \cite{Kitaev2001}. 

The existence of edge states has conventionally been thought to require that perturbations do not close the insulating band (or quasiparticle) gap. This assumption was proven wrong in 1D in a work by Verresen, Jones, and Pollmann \cite{VJP}. These authors showed that exponentially localized edge states may survive at quantum criticality, at the gap-closing quantum phase transition (QPT) between {\em two topologically nontrivial gapped phases}, i.e. phases which both support topological edge states. Earlier works 
\cite{Kestner,Cheng, Fidkowski,Sau,Kraus,Keselman1,Iemini,Lang,Montorsi,Ruhman,Jiang,Zhang,Parker,Keselman2} exploring specific models in 1D had anticipated that distinct topological phases $-$ supporting robust edge states $-$ may in fact form also at quantum critical points (and possibly also in higher dimensions in the presence of additional gapped degrees of freedom \cite{Grover,Scaffidi}). However, why and how this happens was first unveiled in detail by Verresen {\em et al.} \cite{VJP}, providing important intuition. 

Their theory is underpinned by a study of QPTs within the BDI symmetry class of the ten-fold way, where the single-particle Hamiltonians $H$ exhibit spinless time-reversal symmetry,  $THT^{-1} = H,$ particle-hole symmetry $CHC^{-1} = -H$, and chiral symmetry $SHS^{-1} = -H$, where $T, \,C,$ and $S$ are the corresponding first-quantized symmetry operators \cite{Ludwig2016} with $T^2 \!=\! C^2\!=\!1$. The critical point separating the two topologically nontrivial phases is labeled by two numbers: a topological invariant $\nu$, which, if positive, counts the edge states, and the central charge $c$ of the conformal field theory (CFT) which describes the scaling limit of the unperturbed Hamiltonian, with no gapped degrees of freedom present \cite{foot}. The key to the analysis is provided by a meromorphic function which encodes the properties of BDI Hamiltonians, the zeros and poles of which control the values of  $\nu$ and $c$ \cite{VJP}. Given this, Verresen {\em et al.}  argue that two critical Hamiltonians in the BDI class can be smoothly connected (by tuning a control parameter) only if they share the same values of $\nu$ and $c$. In a  follow-up work it is shown that $\nu$ and $c$ can be encoded also by correlation functions of certain nonlocal string order parameters \cite{Jones2019}. The theory was subsequently put into a larger framework of ``symmetry-enriched quantum criticality" \cite{Verresen2019, Vasseur} where the presence of nonlocal symmetry operators implies localized and topologically robust edge modes.  Also, a generalization to all symmetry classes where the topological classification is larger than $\mathds{Z}_2$, independent of dimension, was put forward by Verresen in Ref. \onlinecite{Verresen2020}. 

The case when a lattice system is not strictly translational invariant (with respect to the underlying lattice), but instead has a repeating enlarged unit cell, is briefly touched upon in Ref. \onlinecite{VJPSM}, but then only for the BDI symmetry class, and with no proof of the existence of the critical edge states. In this paper we wish to add to the picture by addressing the unit-cell problem for all three {\em chiral symmetry classes} of noninteracting fermions in 1D, i.e. the symmetry classes in 1D with a topological classification beyond $\mathds{Z}_2$: BDI, CII, and AIII, containing models subject only to perturbations which respect chiral invariance. Note that other symmetry classes in 1D exhibit at most a single topologically nontrivial gapped phase, and therefore, in keeping with Refs.~\onlinecite{VJP, Verresen2020}, do not support critical edge states. One should here recall that an enlarged unit cell implies that the spectrum of the model $-$ be it a band insulator or a mean-field superconductor $-$ displays a multiband structure in the Brillouin zone. Aside from the possible relevance for experiments $-$ including studies of multiband topological nanowires \cite{Lutchyn,Stanescu,Setiawan,Samokhin}, quasi-1D fermionic gases in synthetic gauge fields \cite{Mizushima}, and 1D topological quantum phase transitions out of equilibrium \cite{Mendl,Maslowski} $-$ an analysis of the multiband problem introduces several new facets which may advance our general understanding of topology at quantum criticality. One may here mention that the extended scenario presented in Ref. \onlinecite{Verresen2020} builds on a low-energy representation of an appropriate Hamiltonian and is therefore not directly applicable to multiband systems where higher-energy bands may impact the topological classification. This provides yet another motivation for a more thorough study of the multiband problem, here narrowed to critical chiral phases in 1D.

The paper is organized as follows. In the next section we address the problem of 1D gapless chiral phases in symmetry class AIII. This is the most general case supporting such phases in 1D since only chiral invariance is enforced on the allowed perturbations; time-reversal and particle-hole symmetries, if at all present, are considered as accidental symmetries. As a warm-up, in Sec. II.A we construct the topological invariant for the simple two-band case, borrowing some of the machinery from Ref. \onlinecite{VJP}, but slightly adapted so as to be immediately extendable to the multiband case, as shown in Sec. II.B. In Secs. II.C and II.D the critical edge states are constructed explicitly for the two-band and multiband case respectively. While our approach for the two-band case in Sec. II.C again closely follows Ref. \onlinecite{VJP}, the approach in Sec. II.D is new. Section III contains a test of our approach in Sec. II when applied to symmetry class BDI, here treated as a special case of AIII, where, in addition to chiral symmetry, time-reversal and particle-hole symmetries are also enforced. This section aims to thoroughly compare our results to those in Ref. \onlinecite{VJP} where class BDI is treated explicitly. In Sec. IV we turn to the CII symmetry class, with the same symmetries enforced as for BDI but with $T^2 = C^2 = -1$, thus comprising {\em spinful} free fermion models protected by all three symmetries; chiral, time-reversal, and particle-hole symmetries. {\color{black} For transparency $-$ and also to chisel out the similarities and differences between the treatment of critical BDI models in Ref. \onlinecite{VJP} $-$ we here focus on the case of a CII spinful Majorana chain with four bands.} Section V contains a numerical check of the robustness of critical edge states against uncorrelated disorder, {\color{black} employing the spinful Majorana chain from the previous section as benchmark model.} Section VI, finally, briefly summarizes our work.

\section{symmetry class AIII}
\subsection{Topological invariant for two-band gapless AIII systems in 1D}

A noninteracting fermion system with first-quantized single-particle Hamiltonian $H$ is said to possess chiral symmetry if there exists a unitary operator $S$ such that $SHS^{-1} = -H$. For a lattice system one can then define two effective sublattices associated with this symmetry by using projectors $P_A = (1 + S)/2$ and $P_B = (1 - S)/2$ where $P_{A/B}$ projects states onto the sublattices labelled by $A$ and $B$, respectively. Note that by using this definition, $A$ and $B$ act indeed as sublattices: the terms present in $H$ cannot couple states from the same sublattice due to the restriction imposed by $SH S^{-1} = -H$.    

Now, let us consider a two-band fermionic lattice system from the AIII symmetry class, defined by requiring that all perturbations, and also the Hamiltonian, respect chiral symmetry. By this, any AIII Hamiltonian $H$ can connect only sites from different sublattices, implying the generic expression
\begin{align} 
\begin{split} 
H &= \sum_{j,n} t_n |A,j\rangle \langle B,j + n| + \mbox{H.c.},\\
\end{split}
\label{eq:H1}
\end{align}
where $j$ ($n$) runs from 1 to $N$ (from $-N$ to $N$), with $N$ the number of lattice sites. The two sublattices are denoted by  $A$ and $B$, and $t_n$ are the corresponding {hopping amplitudes, here allowed to be complex but restricted to a finite range, i.e. $t_n = 0$ for large enough $|n|$. Note that $A$ and $B$ denote states which correspond to a pair of internal degrees of freedom. This can be spin, the two sites of a unit cell, or Nambu degrees of freedom.

The Hamiltonian in Eq. (\ref{eq:H1}) is easily diagonalized by a Fourier transformation:
\begin{align} 
\begin{split} 
H = \sum_{k,n} t_n \exp{(i k n)} |A,k\rangle \langle B,k| + \mbox{H.c.},
\end{split}
\label{eq:H2}
\end{align}
with $k = 0, 2 \pi/N, ..., 2 \pi (N-1)/N$. Or simply
\begin{align} 
\begin{split} 
H(k) = \begin{pmatrix} 
0 & f(k) \\
f^\dagger(k) & 0 
\end{pmatrix} ,
\end{split}
\label{eq:H3}
\end{align}
with $f(k) = \sum_{n} t_n \exp{(i  k n)}$ in the basis spanned by $|A,k\rangle, |B,k\rangle$. Importantly, the function $f(k)$ is seen to be in one-to-one correspondence with $H(k)$. It follows from Eq. (\ref{eq:H3}) that the eigenenergies are given by $\epsilon_k =  \pm |f(k)|$ (which means that $f(k) = \epsilon_k e^{i \varphi_k}$ for some $\varphi_k$) and therefore the zeros of $\epsilon_k$ coincide with the zeros of $f(k)$. If nondegenerate, such a zero, $k = k_0$, implies that $\epsilon_k \sim k - k_0$, bringing about a massless relativistic excitation of the critical theory (with the bulk energy gap being closed). When the internal states $A$ and $B$ in  Eq.~(\ref{eq:H1}) are Nambu degrees of freedom, allowing for a nontrivial Majorana representation of the Hamiltonian in Eq. (\ref{eq:H1}), the excitations are those of Majoranas (as in Ref. \onlinecite{VJP}), else they are ordinary fermions. Provided that {\em all} zeros of $f(k)$ are nondegenerate, precluding the appearance of dispersions $\epsilon_k \sim (k-k_0)^m$ with $m \neq 1$ a dynamical exponent, one infers that the effective field theory which describes the critical phase is that of a conformal field theory \cite{VJP}.
 
For gapped systems, with $\epsilon_k\neq 0$,  $f(k)=\epsilon_k \mathrm{e}^{i\varphi_k}$ is a well-defined function on the unit circle with the phase factor $\mathrm{e}^{i\varphi_k}$ prescribing a mapping $S^1 \rightarrow S^1$. By this, the winding number $\nu$ which defines the topological invariant for the 1D AIII symmetry class can be identified with the number of times that $f(k)$ winds around the origin in the complex plane as $k$ is swept through the Brillouin zone (BZ), $k \in [0, 2\pi]$, 
\begin{equation}
    \nu \equiv \frac{1}{2\pi i} \int_{\mathrm{BZ}} dk \,{\color{black} \partial_k \ln(f(k)).} 
\end{equation}
This standard formula for a winding number topological invariant breaks down for {\em gapless} AIII phases, since now $\epsilon_k$ vanishes for at least one value of $k$.  However, as realized by Verresen {\em et al.} for the BDI symmetry class \cite{VJP}, one may circumvent this difficulty by performing an analytic continuation of $f(k)$ to the entire complex plane, i.e., taking $f(k)\rightarrow f(z)$ in such a way that 
\begin{equation} \label {fz}
    f(z)=\sum_{n}t_{n}z^{n}.
\end{equation}
The winding number $\nu$ can now be calculated using Cauchy's argument principle as 
\begin{equation} 
    \nu=N_z-N_p, 
    \label{vnu}
\end{equation}
with $N_z$ the number of zeros of $f(z)$ inside the unit disk, $|z| < 1$, and where $N_p$ is the number of poles (counting multiplicities) also inside the unit disk. Importantly, the right-hand side of (\ref{vnu}) remains well defined also in the gapless case. The only difference to the BDI case is that $t_n$ are now allowed to be complex-valued constants since time-reversal invariance is not enforced for AIII. The quantity on the right-hand side of Eq. (\ref{vnu}) is well defined even for gapless systems and hence can be used for topological distinction of such systems in symmetry class AIII. To readers familiar with the work by Verresen {\em et al.}, let us point out that for the BDI symmetry class (with real-valued hopping amplitudes), a unitary transformation connects the function $f(z)$ in Eq. (\ref{fz}) to its namesake in Ref. \onlinecite{VJP}, constructed using a Majorana representation. For details, see Sec. IV.  

\subsection{Multiband case}

We next consider the extension to the case with $2n$ distinct states per unit cell, with $n>1$. The corresponding single-particle Hamiltonian $H(k)$ in $k$-space will now be represented by a $2n\times 2n$ matrix, implying $2n$ bands in the Brillouin zone.

To set the stage, let us choose a basis for the unit cell that diagonalizes the chiral operator $S$, and call the corresponding basis states  $|X,i\rangle$ (with eigenvalue $+1$) and $|Y,i\rangle$ (with eigenvalue $-1$) where the labels $X$, $Y$ run over the $2n$ different states within a cell and $i$ runs over the cells. We can always go to this basis by a unitary transformation. To satisfy the chiral symmetry condition $SH S^{-1} = -H$ we can have couplings only between states of opposite eigenvalues, in other words, $|X,i\rangle$ can couple only to $|Y,j\rangle$. Therefore, the most general single-particle multiband Hamiltonian can be written as $H = \sum_{X,Y} H_{XY}$, where
\begin{align} 
\begin{split} 
H_{XY}  &= \sum_{i,j} t^{XY}_j |X,i\rangle \langle Y,i + j| + {\mbox H.c.} ,\\
\end{split}
\label{eq:H1_4}
\end{align}
with $i, j$ running over all cell indices. The labels $X$ and $Y$ run over the internal degrees of freedom $A, B, C, D,...$, partitioned according to the evenness and oddness of the corresponding states under chirality, $X \in \{A, B, ...\}$ and $Y \in \{C, D, ...\}$. $t^{XY}_j$, finally, are hopping amplitudes, which, as for the two-band case, are allowed to be complex.  

We now diagonalize {\color{black}$H$} by performing a Fourier transformation, writing
\begin{align} 
\begin{split} 
{\color{black} H = \sum_{X,Y} \sum_{k,j} t^{XY}_j} \exp{(i k j)} |X,k\rangle \langle Y,k| + \mbox{H.c.},
\end{split}
\label{eq:H2_4}
\end{align}
with $k = 0, 2 \pi/N, ..., 2 \pi (N-1)/N$. Reading off the Hamiltonian in $k$-space:
\begin{align} 
\begin{split} 
{\color{black} H(k)} = \begin{pmatrix} 
0 & F(k) \\
F^\dagger(k) & 0 
\end{pmatrix},
\end{split}
\label{eq:H3_4}
\end{align}
with the $n \times n$ matrix $F(k)$ composed of elements $f_{XY}(k) = \sum_{j} t^{XY}_j \exp{(i  k j)}$. For example, with four bands we have 
\begin{align} 
\begin{split} 
F(k) = \begin{pmatrix} 
f_{AC} &  f_{AD} \\
f_{BC} & f_{BD} 
\end{pmatrix}.
\end{split}
\label{eq:H4_4}
\end{align}
Clearly, the eigenenergies are zero if and only if $\det F(k)$ is zero. Therefore, the zeros of {\color{black}$H(k)$} coincide with the zeros of $d(k) = \det F(k)$, with $F(k)$ in one-to-one correspondence with {\color{black}$H(k)$}. 

The winding number $\nu$ for a multiband gapped system is now calculated as
\begin{align} 
\begin{split} 
\nu = \frac{1}{2\pi i} \int dk  \,{\color{black} \partial_k \ln( {\color{black} d(k)}).} 
\end{split}
\label{eq:winding_4}
\end{align}
For the same reason as for the two-band case, this formula becomes inapplicable for gapless phases. To get around this we again use analytic continuation, $d(k) \rightarrow d(z)$, with
\begin{align} \label{dz}
d(z) &= \det \sum_{XY} \sum_j t^{XY}_j z^j |X \rangle \langle Y| \\ \nonumber
&= \det \sum_j T_j z^j, 
\end{align}
where $T_j$ is the corresponding hopping matrix constructed out of the $t^{XY}_j$ coefficients. By employing Cauchy's argument principle, we obtain the same expression for the winding number $\nu$ as in the two-band case, $\nu = N_z -N_p$, now with $N_z$ the number of zeros and $N_p$ the number of poles (including multiplicity) of $d(z)$ inside the unit disk. By construction, this expression for the winding number is valid also for a gapless multiband system in the AIII symmetry class and hence can be used to label its distinct critical phases.

\subsubsection{Extending the unit cell}
Any lattice system invariant under a translation by a unit cell has an equivalent description in which the unit cell has been enlarged, but now with more internal degrees of freedom. In particular, chiral symmetry is preserved under an extension of the unit cell since the new Hamiltonian, written in a basis with the enlarged unit cell, still only couples sites belonging to distinct sublattices. We can enlarge the unit cell until we end up with hopping only between nearest-neighbor cells, i.e. only the matrices $T_{-1}$, $T_{0}$, and $T_{1}$ in Eq. (\ref{dz}) are nonzero (with the row- and column-indices of these hopping matrices now running over all internal states in the extended unit cell). This drastically simplifies the analysis since we are left with only three terms in the summation over $j$ in Eq. (\ref{dz}). As a result, the winding number $\nu = N_z - N_p$ can now be calculated using 
\begin{equation} \label{newdz}
d(z) = \det F(z) = z^{-n} \det (T_{-1} + T_{0}z + T_{1}z^2). 
\end{equation}
The poles in this expression appear only in the prefactor $z^{-n}$ and it follows that we can write $\nu = {\bar N}_z - n$ where ${\bar N}_z$ is the number of zeros inside the unit disk of $\det{\bar F}(z)$, where
\begin{equation} \label{Fz}
\bar{F}(z) = T_{-1} + T_{0}z + T_{1}z^2.
\end{equation} 
Note that the counting of zeros include possible zeros at the origin that will cancel one or several of the poles of $z^{-n}$ in Eq. (\ref{newdz}). This is a useful result that we shall exploit when proving the existence of critical edge states  for a multiband system. Also important to note here that the enlarging of the unit cell does not change the generalized winding number $\nu = N_z - N_p$\cite{VJPSM}. {\color{black} In passing, let us stress that the overbars in the formulas above and in the rest of the paper are used to denote different variables, not complex conjugation.}  

\subsubsection{Short comment}
Any chiral critical noninteracting fermion system in 1D $-$ i.e. any system exhibiting a critical phase and belonging to symmetry class AIII, BDI, or CII $-$ can be classified using the approach above. While we have here exploited only the presence of chiral invariance, assuming symmetry class AIII, the additional symmetries enforced by CII and BDI will only add restrictions on $F(z)$, and hence on $d(z)$ in Eq.(\ref{newdz}). It may also be worth emphasizing that we have not assumed any specific form for the chiral symmetry operator $S$, but only used that $S$ is unitary and that hence there exists a basis which diagonalizes $S$.
\\

\subsection{Edge states in two-band gapless \\ AIII systems in 1D}

We now turn to the demonstration of AIII critical edge states. Our plan of attack for the case of two bands is very similar to that of Verresen {\em et al.} for the BDI symmetry class \onlinecite{VJP} (with substantial changes when we later turn to multiband systems). The goal is to construct $\nu$ linearly independent states (per edge) ``by hand", with the properties that (i) their energies vanish identically for a semi-infinite chain, and (ii) their wave functions decay exponentially as one moves away from the edge, by this establishing a bulk-boundary correspondence \cite{Hasan} for 1D critical two-band systems in the AIII symmetry class. 

\subsubsection{$\nu = N_z - N_p >0$}
For clarity, let us again write the Hamiltonian in Eq. (\ref{eq:H1}), but now explicitly for a semi-infinite chain: 
\begin{align} 
\begin{split} 
H  &= \sum^{\Lambda}_{n= -\Lambda} \sum_{i\ge 1}\, t_n |A,i\rangle \langle B,i + n| + \mbox{H.c.} \\
\end{split}
\label{eq:Edge1}
\end{align}
Here $\pm \Lambda$ are cutoffs beyond which the range of hopping vanishes. The state $|B,i+n\rangle$ is a null vector for $i+n \le 0$.
Introducing a state 
\begin{align} 
\begin{split} 
|\psi_\alpha\rangle = \sum_{i\geq1} \left(a^{(\alpha)}_{i}  | A,i\rangle  + b^{(\alpha)}_{i}| B,i\rangle \right),
\end{split}
\label{eq:Edge2}
\end{align}
this state will represent a zero mode of $H$ if 
\begin{align} 
\begin{split} 
& H|\psi_\alpha\rangle =\\
&= \sum_{n=-\Lambda}^{\Lambda} \sum_{i, j\geq1}\! \left(t_n |A,i\rangle \langle B,i\!+\!n| + \mbox{H.c.}\right)\!(a^{(\alpha)}_{j} |A,j\rangle  + b^{(\alpha)}_{j}|B,j\rangle ) \\
&=  \sum_{n=-\Lambda}^{\Lambda} \sum_{i\geq1} \left(t_n b^{(\alpha)}_{i + n}|A,i\rangle   + t^*_n a^{(\alpha)}_{i} |B,i\!+\!n\rangle \right) = 0, \\
\end{split}
\label{eq:Edge3}
\end{align}
with the sum over unit cells constrained by $i\!+\!n\! \ge \!1$.
This gives us the following conditions on the coefficients which multiply the $|A, i\rangle$ and $|B, i\rangle$ states in Eq. (\ref{eq:Edge3}), call them $C_{A, i}$ and $D_{B,i}$ respectively:  
\begin{subequations}
\begin{align} 
C_{A, i} &\equiv \sum_{{\color{black}m=i-\Lambda}}^{i+\Lambda} t_{m-i} b^{(\alpha)}_{m} = 0 \label{eq:Edge4a}\\
D_{B, i} &\equiv  \sum_{{\color{black}m=i-\Lambda}}^{i+\Lambda} t^*_{i-m} a^{(\alpha)}_{m} = 0, \label{eq:Edge4b}
\end{align}
with $i \ge 1$. 
The idea is now to use the zeros of $f(z)$, Eq.~(\ref{fz}), to prove that the coefficients in Eqs. (\ref{eq:Edge4a}) and (\ref{eq:Edge4b}) can be chosen so as to yield precisely $\nu$ edge states $|\psi_{\alpha}\rangle$ at edge edge of the chain, \, $\alpha = 1,..., \nu$, with the properties (i) and (ii) above. We denote by $z_\alpha$ the largest $\nu$ zeros within the unit disk, with the rest of the zeros denoted $\tilde{z}_s$:  $f(z_\alpha) = 0$ with  $\alpha = 1,..., \nu$, and $f(\tilde{z}_s) = 0$ with  $s= 1,..., N_p$. 
\end{subequations} 
\\

{\em Case $N_p = 0\,$:} Let us first consider the case when there are no poles in $f(z)$. This means that $t_{n<0} = 0$ in Eq. (\ref{fz}) since the corresponding terms are the ones that create poles in $f(z)$. In this case Eqs.~(\ref{eq:Edge4a}) and (\ref{eq:Edge4b}) contain all hopping amplitudes $t_{n}$ present in $f(z)$. It is therefore easy to construct the zero modes by taking $b^{(\alpha)}_{m} = z_\alpha^{m-1}$ and $a^{(\alpha)}_{m} = 0$. It follows that the coefficients reduce to $C_{A, i} =  z_\alpha^{i-1}f(z_\alpha) = 0$ (using that $t_{n< 0} = t_{n> \Lambda} = 0$ in the expression for $f(z_{\alpha}$)) and $D_{B, i} = 0$. The choice $b^{(\alpha)}_{m} = z_\alpha^{m-1}$, rather than the more intuitive $b^{(\alpha)}_{m} = z_\alpha^{m}$, is mandated by the normalizability of a zero mode: With $b^{(\alpha)}_{m} = z_\alpha^{m-1}$, the inner product $\langle \psi_{\alpha} | \psi_{\alpha} \rangle$ is guaranteed to be nonzero,
\begin{eqnarray}
\langle \psi_{\alpha} | \psi_{\alpha} \rangle &=& \sum_{i,j \ge 1} b_i^{(\alpha) \ast} b_j^{(\alpha)}\langle B,i | B, j \rangle =  \sum_{i \ge 1} |z_\alpha|^{2(i-1)} \nonumber \\
& = & \frac{1}{1 - |z_\alpha|^2}, 
\end{eqnarray}
with the second line obtained by summing the geometric series, given that $|z_\alpha| < 1$. Note that we here use a notation where $(z_\alpha)^0 = 1$ also for $z_\alpha = 0$.

The $\nu$ states obtained by inserting  $b^{(\alpha)}_{i} \!= \!z_\alpha^{i-1}$ and $a^{(\alpha)}_{i}\! = \!0$, $\alpha = 1,..., \nu$, into Eq. (\ref{eq:Edge2}) have zero energy by construction, and moreover, they decay with the unit cell index $i$ since $z_\alpha$ is inside the unit disk, implying an exponential decay $\sim \exp(i /\xi_{\alpha})$ with localization length $\xi_{\alpha} = -1/\ln|z_{\alpha}|$. Thus, both conditions (i) and (ii) above are satisfied. Let us note in passing that when a zero $z_{\alpha}$ approaches the unit circle, the localization length is seen to diverge, signaling criticality, with the corresponding edge mode hybridizing with the bulk spectrum, leaving room for a massless bulk excitation at $|z_{\alpha}| = 1$, in accord with the discussion in Sec. II.A.

All cases with distinct $z_\alpha, \alpha = 1,...,N_z$, yield $N_z$ linear independent edge states. But what if there are $m$ degenerate zeros of $f(z)$ (meaning that for $\alpha_1$, ..., $\alpha_m$} the zeros are $z_{\alpha_1} = ... = z_{\alpha_m} = z_\alpha$)? Following Ref. \onlinecite{VJP} we can here take  
\begin{equation} \label{der}
b^{(\alpha_{\ell})}_{m} = \frac{d^{\ell-1} z^{m-1}}{dz^{\ell-1}}|_{z = z_{\alpha}} = \frac{(m-1)!}{(m-\ell)!}z^{m-\ell}_{\alpha}, \ \ \ell= 1,2,...,m, 
\end{equation}
and one verifies that this implies
\begin{equation}
C_{A, i} = z_\alpha^{i-1}\frac{d^{\ell-1}f(z)}{dz^{\ell-1}}|_{z = z_{\alpha}} = 0, \ \ \ell= 1,2,...,m, 
\end{equation}
thus producing $m$ linear independent edge states as required.

Summarizing, when $N_p=0$ we can readily construct $N_z$ edge states with the required properties (i) and (ii), valid also when the system is critical. Note that the edge states thus obtained are nonzero only on sublattice $B$ since we have chosen $a^{(\alpha)}_{m} = 0$. One should here note that given a basis in which the chiral symmetry operator $S$ is diagonal, zero-energy edge states of {\em any} chiral-symmetric model necessarily have support on only one sublattice of a bipartite lattice \cite{AsbothBook}.
\\

{\em Case $N_p \neq 0\,$:} In this case we look for $b^{(\alpha)}_{m} = z_\alpha^{m} + \sum_{s= 1}^{N_p} \lambda^{(\alpha)}_{s} \tilde{z}_s^{m}$, with $\{\lambda^{(\alpha)}_{s}\}$ complex numbers to be determined. As before we take all $a^{(\alpha)}_{m} = 0$. We know that $t_{n < -N_p} = 0$ since otherwise the multiplicity of the pole at the origin would be larger than $N_p$. It follows from Eq. (\ref{eq:Edge4a}) that the condition $C_{A, i > N_p} = 0$ is trivially satisfied. It remains to find a set of constants $\lambda^{(\alpha)}_s$ that will make the rest of the expansion coefficients $C_{A, 1\leq i \leq N_p}$ vanish identically. Again invoking Eq. (\ref{eq:Edge4a}), these constants must satisfy the equations 
\begin{align} 
\begin{split} 
A_{i s}\lambda_s^{(\alpha)} = - \sum_{{\color{black}m =i-\Lambda}}^{i+\Lambda}t_{m-i}z_{\alpha}^m, \ \ i=1,2,...,N_p,
\end{split}
\label{eq:Edge5}
\end{align}
where $A_{is} = \sum_{{\color{black}m=i-\Lambda}}^{i+\Lambda} t_{m-i} \tilde{z}_s^{m}$ and summation over $s$ is implied. These equations are equivalent to a matrix equation of the type $A \lambda = b$, with $A$ an $N_p \times N_p$ matrix with elements $A_{is}$. As follows from the discussion in Ref. \onlinecite{VJPSM}, provided that $z_{\alpha}$ is nondegenerate, a row reduction turns $A$ into an invertible Vandermonde matrix, implying the existence of unique solutions for the $\lambda^{(\alpha)}_s$. 

If there are degenerate zeros of $f(z)$ one can employ the same device as in the case with degeneracies when $N_p=0$, working with derivatives instead; cf. Eq. (\ref{der}). However, we shall not go into details here.

\subsubsection{$\nu = N_z - N_p <0$}

Turning to the case of negative winding numbers, we cannot directly employ the elegant inversion argument used in Ref. \onlinecite{VJP} since in our case, with complex hopping amplitudes allowed in the AIII symmetry class, inversion symmetry is in general broken. However, the necessary modification of the inversion argument is minor, and essentially comes down to doing some ``relabeling" in the preceding equations for positive winding numbers. 

We start off with the same Hamiltonian $H$ as before, Eq. (\ref{eq:Edge1}), 
and interchange $A$ with $B$, and $n$ with $-n$. We thus obtain
\begin{align} 
\begin{split} 
H^\prime  &=  \sum_{n=-\Lambda}^{\Lambda} \sum_{i\ge1}\, t^\prime_n|A, i-n\rangle \langle B,i| + \mbox{H.c.},\\
\end{split}
\label{eq:Edge7}
\end{align}
with $t^\prime_n = (t_{-n})^*$  and with $|A,i - n\rangle$ a null vector for $i-n \le 0$. Note that this is just a rewriting of the Hamiltonian and hence the edge states of $H$ and $H^\prime$ must be the same. Let us focus on $H^\prime$, in one-to-one correspondence with  $f^\prime(z) = \sum_n(t_{-n})^* z^n = f^*(1/z^*)$ (where the summation is over $n \in [-\Lambda, \Lambda]$).
Now assume that $f(z)$ has $N=N_\text{Z} + N_\text{C} + N_\text{O}$ total number of zeros, with $N_\text{C} \,(N_\text{O})$ the number of zeros on (outside) the unit circle. We can then write
\begin{equation} \label{decomp}
f^{\prime}(z) = \frac{1}{z^{-N_\text{P}}}\prod_{i=1}^N(\frac{1}{z}-z_i^*) = \frac{(-1)^N}{z^{N-N_\text{P}}}\prod_{i=1}^{N}(z-\frac{1}{z_i^*})z_i^*,
\end{equation}
with, as before, $N_\text{P}$ the multiplicity of the pole of $f(z)$ at the origin. It follows that the multiplicity of the pole of $f^{\prime}(z)$ is $N_P^{\prime} = N - N_\text{P}$. Moreover, by the inversion $z \rightarrow 1/z$, $f'(z)$ must have the same number of zeros inside the unit circle as $f(z)$ has outside the unit circle, that is, $N_\text{Z}^{\prime} = N_\text{O}$. The winding number $\nu^{\prime}$ for the inverted system is now easily calculated as
\begin{eqnarray} \label{inverted}
\nu^{\prime} & = &  N_\text{Z}^{\prime} - N_\text{P}^{\prime} = N_\text{O} - (N - N_\text{P}) \nonumber \\
& = & N_\text{O} - (N_\text{Z} + N_\text{O} + N_\text{C} - N_\text{P}) = - \nu - N_\text{C}.
\end{eqnarray} 

Given this result we can construct the edge states for $\nu < 0$ in exact analogy to the case $\nu > 0$, using the zeros of $f^\prime(z)$. Acting with $H^{\prime}$, Eq. (\ref{eq:Edge7}), on a state $|\psi_\alpha\rangle = \sum_{i\geq1} (a^{(\alpha)}_{i}  | A,i\rangle  + b^{(\alpha)}_{i}| B,i\rangle)$, the condition that this state is a zero mode of $H^{\prime}$ takes the form    
\begin{align} 
\begin{split} 
& H^\prime|\psi_\alpha\rangle =\\
&=  \sum_{n=-\Lambda}^{\Lambda}\sum_{i\ge 1} \left(t^\prime_n b^{(\alpha)}_{i}|A,i - n\rangle   + (t^\prime_n)^* a^{(\alpha)}_{i - n} |B,i\rangle \right) = 0, \\
\end{split}
\label{eq:Edge10}
\end{align}
with the sum over unit cells constrained by $i-n \ge 1$. It follows that the coefficients $C^{\prime}_{A,i}$ and $C^{\prime}_{B,i}$ which multiply the $|A,i\rangle$ and $|B,i\rangle$ states respectively must satisfy
\begin{subequations}
\begin{align} 
C^{\prime}_{A, i} &=  \sum_{{\color{black}m=i-\Lambda}}^{i+\Lambda} t^\prime_{m-i} b^{(\alpha)}_{m} = 0 \label{eq:Edge11a}\\
D^{\prime}_{B, i} &=  \sum_{{\color{black}m=i-\Lambda}}^{i+\Lambda} (t^\prime_{i-m})^* a^{(\alpha)}_{m} = 0 \label{eq:Edge11b}
\end{align}
\end{subequations} 
with $i \ge 1$. These are the same equations as for the case with $\nu > 0$, Eqs. (\ref{eq:Edge4a}) and (\ref{eq:Edge4b}), and hence we can construct the edge states in exactly the same manner, but now using the zeros of $f^\prime(z)$ that will give us $\nu^\prime = - \nu - N_\text{C}$ edge states. It may be worth pointing out that these states live on sublattice $A$, reflecting the fact that we interchanged $A$ and $B$ when rewriting the Hamiltonian, replacing $H$ by $H^{\prime}$. 

\subsection{Edge states in $\boldsymbol{2n}$-band gapless \\ AIII systems in 1D}

We now proceed to generalize our finding from the previous section and establish a bulk-boundary correspondence for 1D critical {\em multiband} systems in the AIII symmetry class, i.e. systems with $2n$ bands  (with $n \ge 1$, treating two bands as a special case). As for the simple two-band case, our objective is to construct $\nu$ linearly independent states per edge with the properties that (i) their energies vanish identically for a semi-infinite chain, and (ii) their wave functions decay exponentially as one moves away from the edge. Here we shall take the unit cell large enough so that only the hopping matrices $T_{0}, T_{\pm 1}$ in Eq.~(\ref{dz}) are nonzero. This is a convenient choice which simplifies the analysis, and which does not impact the result for the number of edge states.    

\subsubsection{$\nu = N_z - N_p >0$}

For transparency, let us begin by explicitly writing down the Hamiltonian in Eq. (\ref{eq:H1_4}) on a semi-infinite lattice, having chosen a sufficiently large unit cell so that there are only three sets of nonzero hopping amplitudes $t_j^{XY}$, with $j=0, \pm 1$:
\begin{align} 
\begin{split} 
H  &=\!\sum_{-1\le j \le 1} \sum_{i\ge 1} t^{XY}_j |X,i\rangle \langle Y,i + j| + \mbox{H.c.},\\
\end{split}
\label{eq:Edge1_4}
\end{align}
where $|Y,0\rangle$ is a null vector. As spelled out in Sec. II.B, the labels $X$ and $Y$ run over the $2n$ different states within a unit cell, $X \!\in\!\{A, B, ...\}$ on one sublattice and $Y\! \in\!\{C, D, ...\}$  on the other, with $X$ and $Y$ implicitly summed over in Eq. (\ref{eq:Edge1_4}). From now on, all occurrences of repeated indices $X$ and $Y$ are summed over. Next, we write a general expression for a  multiband state, 
\begin{align} 
\begin{split}  
|\psi_\alpha\rangle = \sum_{i\geq1} \left(a^{(\alpha)}_{i,X} |X,i\rangle  + b^{(\alpha)}_{i,Y}|Y,i\rangle \right).
\end{split}
\label{eq:Edge2_4}
\end{align}
This state is a zero mode of $H$ if
\begin{eqnarray}
H|\psi_\alpha\rangle &\!=\!& \! \! \sum_{-1\le j \le 1} \,\sum_{i\geq 1} (t^{XY}_j b^{(\alpha)}_{i \!+\! j,Y}|X,i\rangle  \!+\! (t^{XY}_j)^* a^{(\alpha)}_{i,X} |Y,i\!+\!j\rangle) \nonumber \\ 
&= &0. 
\label{eq:Edge3_4}
\end{eqnarray}
Similarly to the two-band case in Sec. II.C, Eqs. (\ref{eq:Edge4a}) and (\ref{eq:Edge4b}), this gives us the following constraints on the coefficients which multiply the $|X,i\rangle$ and $|Y,i\rangle$ states:
\begin{subequations}
\begin{align} 
C_{X, i} &=\sum_{{\color{black}m=i-1}}^{i+1} t^{XY}_{m-i} b^{(\alpha)}_{m,Y} = 0,\\
D_{Y, i} &= \sum_{{\color{black}m=i-1}}^{i+1} (t^{XY}_{i-m})^* a^{(\alpha)}_{m,X} = 0,
\end{align}
\label{eq:Edge4_4}
\end{subequations} 
{\color{black} with $i\ge1$.} Using matrix notation,
\begin{subequations}
\begin{align} 
\boldsymbol{C}_{i} &= \sum_{{\color{black}m=i-1}}^{i+1} T_{m-i} \boldsymbol{b}^{(\alpha)}_{m} = 0 \label{eq:Edge5_4a}\\
\boldsymbol{D}_{i} &= \sum_{{\color{black}m=i-1}}^{i+1} T^*_{i-m}  \boldsymbol{a}^{(\alpha)}_{m} = 0 \label{eq:Edge5_4b}
\end{align}
\end{subequations} 
where $T_{m-i}$ is the hopping matrix constructed out of the amplitudes $t^{XY}_{i}$ (and similarly for $T^*_{i-m}$), and where $\boldsymbol{a}^{(\alpha)}_{m}$ and $\boldsymbol{b}^{(\alpha)}_{m}$ are vectors with elements $a^{(\alpha)}_{m,X}$ and $b^{(\alpha)}_{m,Y}$ respectively.
\\
\\
{\em Case $T_{-1} \!= \!0$:} 
For this case, with only $T_0$ and $T_1$ being nonzero, it is straightforward to construct the zero modes. Recall from Sec. II.B that the winding number $\nu$ can be calculated from $F(z) = T_{-1}z^{-1} + T_{0} + T_{1}z$ as $\nu = N_z - N_p$ where $N_z$ ($N_p$) is the number of zeros (poles) of $\det F(z)$. When $T_{-1} = 0$ there are no poles, and the expression for the winding number simplifies to $\nu = N_z$. Let us take a more pragmatic route here and simplify the computations by assuming that every zero of $\det F(z)$ is nondegenerate. This is motivated by the fact that since a degeneracy will split unless the hopping amplitudes are fine-tuned, any generic experimental uncertainty will wash away the degeneracies, making the degenerate case irrelevant for applications.

Any zero of $\det F(z)$, denoted by $z_\alpha$ with $\alpha = 1,...,N_z$, guarantees that there exists a nonzero eigenvector $\bar{\boldsymbol{b}}^{(\alpha)}$ satisfying $F(z_\alpha) \bar{\boldsymbol{b}}^{(\alpha)} = 0$. By taking $\boldsymbol{b}^{(\alpha)}_{m} = z_\alpha^{m - 1}\bar{\boldsymbol{b}}^{(\alpha)}$ in Eq. (\ref{eq:Edge5_4a}) it is easy to verify that the coefficient $\boldsymbol{C}_{i}$ reduces to $\boldsymbol{C}_{i} =  z_\alpha^{i-1} F(z_\alpha) \bar{\boldsymbol{b}}^{(\alpha)} = 0$. Choosing $\boldsymbol{a}^{(\alpha)}_{m} = 0$ in Eq. (\ref{eq:Edge5_4b}) one infers the existence of $\nu$ zero-energy states, living on one of the sublattices. 

In exact analogy to the two-band case, the states obtained via the construction above $-$ inserting $b_{i,Y}^{(\alpha)}= z_{\alpha}^{i-1}\bar{b}_Y^{(\alpha)}$ and $a_{i,X}^{(\alpha)}=0$ into Eq. (\ref{eq:Edge2_4}), with $a_{i,X}^{(\alpha)}$ and $b_{i,Y}^{(\alpha)}$ the elements of $\boldsymbol{a}^{(\alpha)}_{i}$ and $\boldsymbol{b}^{(\alpha)}_{i}$ $-$ decay exponentially as $\exp(i /\xi_{\alpha})$ with localization length $\xi_{\alpha} = -1/\ln|z_{\alpha}|$, $|z_{\alpha}|<1$. Thus, also condition (ii) above is satisfied.    
\\
\\
\\ 
{\em Case $T_{-1} \!\neq \!0$:}  Let us denote by $z_\beta$ (with $|z_\beta|\!<\!1$) and by~$\bar{\boldsymbol{b}}^{(\beta)}$ all solutions to 
\begin{align} 
\begin{split}
\bar{F}(z_\beta)\bar{\boldsymbol{b}}^{(\beta)} = (T_{-1} + T_{0}z_\beta + T_{1}z_\beta^2)\bar{\boldsymbol{b}}^{(\beta)}= 0.
\end{split}
\label{eq:Edge_mult}
\end{align}
Assuming that all zeros are nondegenerate (cf. the discussion above when $T_{-1} = 0$), their total number is $\bar{N}_z$. {\color{black} The pairs ($z_\beta$; $\bar{\boldsymbol{b}}^{(\beta)}$) define vectors} $\boldsymbol{b}^{(\beta)}_{m} = z_{\beta}^{m-1}\bar{\boldsymbol{b}}^{(\beta)}$ (taking $\boldsymbol{a}^{(\beta)}_{m} = 0$) that {\color{black} are all linearly-independent among each other because they decay at distinct rates $z_\beta^{m-1}$. Given this, we now aim to construct independent edge states out of} $\boldsymbol{b}^{(\beta)}_{m}$. {\color{black} Thus, we look at superpositions  $\boldsymbol{b}^{(\alpha)}_{m} = \sum_{\beta = 1}^{\bar{N}_z}\lambda_{\alpha \beta} \boldsymbol{b}^{(\beta)}_{m}$ and notice that for each such superposition} the coefficients $\boldsymbol{C}_{i\ge 2} = T_{-1}\boldsymbol{b}^{(\alpha)}_{i-1} + {\color{black} T_0 \boldsymbol{b}^{(\alpha)}_{i} + T_1 \boldsymbol{b}^{(\alpha)}_{i+1}}$ in Eq.~(\ref{eq:Edge5_4a}) vanish identically. The existence of zero-energy states  then hinges on the possibility to construct {\color{black} independent vectors $\boldsymbol{b}^{(\alpha)}_{m}$} such that $\boldsymbol{C}_{1}  = \sum_{\beta = 1}^{\bar{N}_z}\lambda_{\alpha \beta}  \boldsymbol{v}_\beta  = 0$ with $\boldsymbol{v}_\beta = T_0 \boldsymbol{b}^{(\beta)}_{1}  + T_1 \boldsymbol{b}^{(\beta)}_{2}$. 

{\color{black} To take on this task, from all $\{\boldsymbol{v}_\beta\}$ we first select a set of basis vectors $\{ \boldsymbol{v}^\text{base}_\gamma\}$} spanning the space of $\{\boldsymbol{v}_\beta\}$. Their number cannot exceed $n$ (dimension of the vectors $\boldsymbol{v}_\beta$). {\color{black} Let us refer to the rest of the vectors in $\{\boldsymbol{v}_\beta\}$ as $\{\boldsymbol{v}^\text{nonbase}_\delta\}$. Each non-basis vector $\boldsymbol{v}^\text{nonbase}_\delta$ may then be decomposed as $\boldsymbol{v}^\text{nonbase}_\delta = \sum_{\gamma}x_{\delta \gamma}  \boldsymbol{v}^\text{base}_\gamma$ with some coefficients $x_{\delta \gamma}$. It follows that for each $\delta$ we can then construct a state from $\boldsymbol{b}^{(\alpha)}_{m} = \boldsymbol{b}^{(\delta)}_{m} - \sum_{\gamma}x_{\delta \gamma} \boldsymbol{b}^{(\gamma)}_{m}$ which trivially fulfills the condition $\boldsymbol{C}_{1}  = 0$.  Importantly, the states thus obtained are linearly independent since they are each tied to a particular component $\boldsymbol{b}^{(\delta)}_{m}$, with each $\boldsymbol{b}^{(\delta)}_{m}$ having a unique decay rate $z_\delta^{m-1}$.

Following the protocol above, the vectors $\boldsymbol{b}^{(\alpha)}_{m} = \boldsymbol{b}^{(\delta)}_{m} - \sum_{\gamma}x_{\delta \gamma} \boldsymbol{b}^{(\gamma)}_{m}$} are seen to define at least $\nu = \bar{N}_z - n = N_z - N_p$ independent edge states. This concludes the analysis for this case.
   
\subsubsection{$\nu = N_z - N_p <0$}

We approach this case in the same way as when we analyzed negative winding numbers for two bands in Sec. II.B, essentially relying only on a ``relabelling" of the equations derived for positive winding numbers. 

By interchanging the sublattice indices $X$ and $Y$ and flipping the sign of the index $j$ which labels the hopping amplitudes $t_j^{XY}$ in Eq. (\ref{eq:Edge1_4}), the multiband Hamiltonian $H$ in the same equation takes the form 
\begin{align} 
\begin{split} 
H^\prime  &= \sum_{-1 \le j \le 1} \sum_{i\ge 1} (t^\prime_j)^{XY} |X, i-j\rangle \langle Y,i| + \mbox{H.c.} \\
\end{split}
\label{eq:Edge10_4}
\end{align}
with $(t^\prime_j)^{XY} = (t^{XY}_{-j})^*$ and where $|X,0\rangle$ is a null state. Recall that repeated indices $X$ and $Y$ are always summed over. The function $F(z) = T_{-1}z^{-1} + T_{0} + T_{1}z$, associated to~$H$, now gets replaced by a new function $F^\prime(z)$, with the property that $\det F^\prime(z) = (\det F(1/z^*))^*$. By exactly the same argument as for the two-band case, it follows that if $\det F(z)$ calls for $\nu < 0$ then $\det F^\prime(z)$ entails $\nu^\prime = - \nu - {N}_\text{C}$ with ${N}_\text{C}$ the number of zeros of $\det F(z)$ on the unit circle.

The construction of the edge states is simple, and can be carried out with the case $\nu>0$ as a template. To see how, we act with $H^{\prime}$ on the multiband state $|\psi_\alpha\rangle$ in Eq. (\ref{eq:Edge2_4}), reading off the condition that $|\psi_\alpha\rangle$ is a zero-energy state:  
\begin{multline}
H^\prime|\psi_\alpha\rangle = \sum_{-1 \le j \le 1} \sum_{i\geq1}  \Big( (t^\prime_j)^{XY} b^{(\alpha)}_{i, Y}|X,i - j\rangle \nonumber \\
+ ((t^\prime_j)^{XY})^* a^{(\alpha)}_{i - j, X} |Y,i\rangle \Big) = 0, 
\label{eq:Edge13_4}
\end{multline}
where $a_0=0$. This equation implies the following constraints on the coefficients (for $i \geq 1$):
\begin{subequations}
\begin{align} 
C_{X, i} &= \sum_{{\color{black}m=i-1}}^{i+1} (t^\prime_{m-i})^{XY} b^{(\alpha)}_{m,Y} = 0\\
D_{Y, i} &= \sum_{{\color{black}m=i-1}}^{i+1} ((t^\prime_{i-m})^{XY})^* a^{(\alpha)}_{m,X} = 0
\end{align}
\label{eq:Edge14_4}
\end{subequations} 
or, using matrix notation,
\begin{subequations}
\begin{align} 
\boldsymbol{C}_{i} &= \sum_{{\color{black}m=i-1}}^{i+1} T^\prime_{m-i} \boldsymbol{b}^{(\alpha)}_{m} = 0\\
\boldsymbol{D}_{i} &=  \sum_{{\color{black}m=i-1}}^{i+1} (T^\prime_{i-m})^*  \boldsymbol{a}^{(\alpha)}_{m} = 0
\end{align}
\label{eq:Edge15_4}
\end{subequations} 
These are the same equations as (\ref{eq:Edge5_4a}) and (\ref{eq:Edge5_4b}) for the case with $\nu > 0$, but with $T$ replaced by $T'$. Thus, we can construct the zero-energy edge states in exactly the same manner as for $\nu>0$, but now using the zeros of $\det F^\prime(z)$ that will give us $\nu^\prime = - \nu - N_\text{C}$ such states. 

\section{Symmetry class BDI}
\subsection{Two-band gapless BDI systems in 1D: topological invariant and edge states}

The problem of critical edge states in the BDI symmetry class of 1D models was solved by Verresen {\em et al.}\cite{VJP} for the case of two bands. Since any model which belongs to the BDI symmetry class is also contained in AIII $-$ the symmetry class which we analyzed in the previous section, including the multiband case $-$ one may think that there is not much to add. Still, it is instructive to revisit the problem to see how the BDI formalism in Ref. \onlinecite{VJP} fits into the scheme presented here, taking into account the symmetries which mark out the BDI class. Beyond chiral invariance, intrinsic also to AIII, these are time-reversal symmetry, $THT^{-1} = H$, and particle-hole symmetry, $CHC^{-1} = -H$, with  $T^2 \!=\! C^2\!=\!1$. To avoid any misconception, it is important to stress that these symmetries are to be considered as {\em accidental} when a Hamiltonian $H$ is placed in the AIII symmetry class, while being {\em enforced} on any perturbation of the same Hamiltonian when placed in BDI. In particular, this implies that the symmetry-protection of edge states is stronger if $H$ is placed in AIII rather than in BDI. 

Verresen {\em et al.} \cite{VJP} performed their analysis on a representative BDI Hamiltonian, $H_\text{BDI}$, describing a spinless superconducting wire. In real space, assuming translational invariance on a chain with unit cells labeled by $j$, the second-quantized Hamiltonian can be written as  
${\cal H}_\text{BDI} = \sum_n t_n {\cal H}_n$, where
\begin{align} 
\begin{split} 
{\cal H}_n &=   \frac{i}{2} \sum_j \tilde{\gamma}_j \gamma_{j + n}\\
&= -\frac{1}{2} \sum_j (c^\dagger_j c^\dagger_{j + n} + c^\dagger_j c_{j + n} - c_j c^\dagger_{j + n} - c_j c_{j + n}). \\
\end{split}
\label{eq:H1_BD1}
\end{align}
Here $\gamma_j \!= \!\frac{1}{2}(c_j^\dagger + c_j)$ and $\tilde{\gamma}_j \!= \! \frac{i}{2}(c_j^\dagger - c_j)$ are Majorana modes, with $c_j$ and $c_j^\dagger$ fermion operators. Note that time-reversal symmetry precludes terms of the form $i\gamma_j \gamma_\ell$ and $i \tilde{\gamma}_j\tilde{\gamma}_\ell$ as well as complex hopping amplitudes in the decomposition of $H_\text{BDI}$. As implied by the notation, each unit cell with index $j$ contains two Majorana modes, $\gamma_j$ and $\tilde{\gamma}_j$.

To make contact with our generic first-quantized AIII Hamiltonian in Eq. (\ref{eq:H1}) we can introduce Nambu spinors  
$\Psi^{\dagger} = (c_1^\dagger ... c_N^\dagger \, c_1 ... c_N)$ and extract the corresponding Bogoliubov-de-Gennes {\color{black} (BdG)} Hamiltonian, call it $H_\text{BdG}$, from                    
${\cal H}_\text{BDI} = \Psi^\dagger H_\text{BdG} \Psi$. By this, one obtains
\begin{equation} \label{BdG}
H_\text{BdG} =   -\frac{1}{4} \sum_{j,n} t_n \left(\tau_z \otimes |j\rangle \langle j + n| + i\tau_y \otimes |j\rangle \langle j + n|\right) + \mbox{H.c.}, 
\end{equation}
\\
which acts on basis states $|\tau\rangle\otimes|j\rangle$, with $|\tau=1\rangle = (1 0)^T$ for a particle state and $|\tau\!=\!-\!1\rangle = (0 1)^T$ for a hole state, and where                          
the state $|j\rangle = (0,...., 1, ....0)^T$ corresponds to the $j$:th site of the chain. $\tau_y$ and $\tau_z$ are Pauli matrices. 

Particle-hole symmetry is built into $H_\text{BdG}$, with $C=\tau_x{\cal K}$ (where $\tau_x$ is the Pauli $x$-matrix and ${\cal K}$ the complex-conjugation operator), as is also           time-reversal symmetry with $T={\cal K}$, with both operators here written in a $k$-space representation. Here $T^2 = C^2 = 1$ as must be since the model is spinless. The chiral symmetry operator $S$ is given by $\tau_x$ in the chosen basis. We can make a rotation to a ``chiral basis" where $S$ is diagonal by simply acting with $U = (\mathbb{1} + \tau_y)/\sqrt{2}$ on the basis states. As a result, $\tau_z \rightarrow \tau_x$ and $\tau_y \rightarrow \tau_y$ and therefore $H_\text{BdG}  \rightarrow  -\frac{1}{4} \sum_{j,n} t_n (\tau_x \otimes |j\rangle \langle j + n| + i\tau_y \otimes |j\rangle \langle j + n|) 
+ \mbox{H.c.}$, now with chiral symmetry operator 
$S = \tau_z$. Rewriting $H_\text{BdG}$ in terms of the eigenstates of $\tau_z$, call them $|A\rangle$ and $|B\rangle$, one obtains $H_\text{BdG} = -\frac{1}{2} \sum_{j,n} t_n |A,j\rangle \langle B,j + n|) + \mbox{H.c.}$, of the very same form as our Eq. (\ref{eq:H1}) (up to an immaterial prefactor of $-1/2$). It follows that the $f(z)$ function derived for the BDI symmetry class in Ref. \onlinecite{VJP} is given by the same expression as in our Eq. (\ref{fz}), with the crucial difference that the parameters $t_n$ are now constrained to be real. According to the fundamental theorem of algebra, this restricts the zeros of $f(z)$ to be real or to come in complex-conjugate pairs. Apart from this, the overall picture does not change. As worked out by Verresen {\em et al.} \cite{VJP}, the number of topological edge states is determined by the zeros and poles inside the unit disk of the $f(z)$ function, just as for AIII.

\subsection{$\boldsymbol{2n}$-band gapless BDI systems in 1D:\\ topological invariant and edge states}

In Ref. \onlinecite{VJPSM}, Verresen {\em et al.} presents a formal proof that the topological invariant for a critical translational invariant system in the BDI symmetry class is independent of the choice of unit cell. While intuitively clear that this must be the case, the proof primarily serves as a consistency check of the construction of an $f(z)$ function when the unit cell is enlarged. If we restrict the generic AIII multiband Hamiltonian in Eq. (\ref{eq:H1}) to BDI by requiring invariance under time-reversal and particle-hole symmetry where $T = \mathbb{1} \otimes \mathbb{1} \mathcal{K}$ and $C = \tau_x  \otimes \mathbb{1} \mathcal{K}$ in a $k$-space representation our $d(z)$ function in Eq. (\ref{dz}), now restricted by these symmetries, can be shown to be identical to that of Ref. \onlinecite{VJPSM}, denoted by $f(z)$ in Eq. (S8) of that same reference. Note that the unit matrix multiplying ${\cal K}$ in the expressions for $T$ and $C$ acts in the $n$-dimensional space of internal states contained in the unit cell (not counting the Nambu degree of freedom with one particle- and one hole-state). The necessary calculation is straightforward but long-winded, and for this reason we omit it here. Moreover, the result is entirely expected: Our construction of critical edge states for multiband AIII models ensures the existence of exponentially localized zero-energy edge states also for critical multiband models in the BDI symmetry class, with the same counting as before, $\nu = N_z- N_p$.

\section{Symmetry class CII}
A model which is placed in symmetry class CII is protected by all three symmetries $S, \,T,$ and $C$ of the tenfold way \cite{Chiu2016}, analogous to BDI, but now with $T^2 = C^2 = -1$. Thus, this is the appropriate symmetry class to use for chiral-invariant {\em spinful} fermion systems with topological edge states protected by time-reversal and particle-hole symmetry. We address this problem similar to how we approached the BDI class in the previous section, where we analyzed how the functions $f(z)$ (for two bands, Eq. (\ref{fz})) and $d(z)$ (for the general multiband case, Eq. (\ref{dz})) get restricted by the additional symmetries $T$ and $C$. Here, we have to proceed with some care. 

First, we cannot satisfy the constraints $T^2 = C^2 = -1$ for a two-band model. The reason is simple. For these constraints to be satisfied, the symmetry operators $T$ and $C$, both being anti-unitary, have to be represented by a product of a Hermitian (and unitary) matrix with purely imaginary elements and the complex conjugation operator ${\cal K}$. There is only {\em one} such $2\times 2$ matrix, the Pauli $y$-matrix, and hence, with $T$ and $C$ being distinct operators, there is no faithful representation adequate for two bands. 

Secondly, turning to more than two bands, one still has to be wary: There are multiple choices for the representation of the symmetries and they may influence the $d(z)$ function in Eq. (\ref{dz}) differently. Here, for ease and clarity, we limit our discussion to the simplest multiband case, $2n=4$, with conventional realizations of the time-reversal and particle-hole  symmetries; see below.

\subsection{Topological invariant for four-band gapless \\ CII systems in 1D}
For the purpose of connecting the representative four-band CII models to possible physical realizations (for examples, see e.g. Refs. \onlinecite{Zhao2014,Prakash2018}), we shall begin by identifying the possible varieties of {\em spinful} superconducting chains that satisfy the required symmetries. The case of four bands is easy to picture, thinking in terms of a Bogoliubov-de-Gennes construction using Nambu spinors as in our discussion of the BDI symmetry class, but now with an added spin-1/2 degree of freedom which doubles the number of bands. We take the time-reversal symmetry operator to be of conventional type for spin-1/2 systems, $T=\mathbb{1} \otimes \sigma_y \mathcal{K}$ (in a $k$-space representation), with the first (second) factor acting in the particle-hole (spin) space, and with ${\cal K}$ acting in both, {\color{black} yielding $T^2 = -1$.} The particle-hole symmetry is decoupled from the spin degree of freedom and therefore its symmetry operator has to be $C = \tau_y \otimes \mathbb{1}\mathcal{K}$ (in the same representation) in order to fulfill the requirement $C^2 = -1$. Here $\tau_y$ is also a Pauli $y$-matrix, but acting in the particle-hole space (cf. the text after Eq. (\ref{BdG})). The chiral symmetry operator $S$ is fixed by $S=T^{-1}C$. In the following we shall explore how the restrictions implied by the $T$ and $C$ symmetries map out the possible spinful chains belonging to the CII class. With an eye to parallels with the BDI symmetry class as studied in Ref. \onlinecite{VJP}, we will work in a representation with Majorana modes. Different from Ref. \onlinecite{VJP}, the Majoranas will now carry spin-1/2.   

\subsubsection{{\color{black} Spinful Majorana chains}}

To construct the possible CII Majorana chains we must analyze what types of spinful Majorana bilinears that survive the restrictions from time-reversal and particle-hole symmetry. Our strategy is to first construct all real-space first-quantized Hamiltonian matrices $H$ allowed by these symmetries and then use a Nambu representation to go to second-quantization, and from there, to a representation in spinful Majoranas, {\color{black} defined by $\gamma_{j,\sigma} = \frac{1}{2}(c^\dagger_{j,\sigma} + c_{j,\sigma})$
and $\tilde{\gamma}_{j,\sigma} =\frac{i}{2}(c^\dagger_{j,\sigma} - c_{j,\sigma})$, with $c^{\dagger}_{j,\sigma}$ and $c_{j,\sigma}$ fermion operators at lattice site $j$ with spin $\sigma = \uparrow, \downarrow$. As an outcome we obtain} the spinful Majorana chains
\begin{align} 
\begin{split} 
{\cal H}^\text{CII}  =  \sum_{n=-\Lambda}^{\Lambda}\! (t_n {\cal H}^\text{CII}_{n}  + \bar{t}_n \bar{\cal H}^\text{CII}_{n}),
\end{split}
\label{eq:H_CII10}
\end{align}
with 
\begin{align} 
\begin{split} 
&{\cal H}^\text{CII}_{n}  =  i \sum_{j, \sigma} \tilde{\gamma}_{j, \sigma} \gamma_{j + n, \sigma}
\end{split}
\label{eq:H_CII11}
\end{align}
and
\begin{align} 
\begin{split} 
&\bar{\cal H}^\text{CII}_{n}  =  i \sum_{j} (\tilde{\gamma}_{j, \uparrow} \gamma_{j + n, \downarrow} -  \tilde{\gamma}_{j, \downarrow}\gamma_{j + n, \uparrow})\\
\end{split}
\label{eq:H_CII12}
\end{align} 
serving as bases for all {\em spin-preserving} and {\em spin-flipping} Majorana chains, respectively. The amplitudes $t_n $ and $\bar{t}_n$ are real, with $[-\Lambda, \Lambda]$ the range of couplings between the Majoranas. The associated hopping matrix in spin space, constructed as in Sec. II.B, is readily read off from Eqs. (\ref{eq:H_CII10}) - (\ref{eq:H_CII12}): 
\begin{align} 
\begin{split} 
T_n = \begin{pmatrix} 
t_n &  \bar{t}_n \\
-\bar{t}_n & t_n 
\end{pmatrix}.
\end{split}
\label{eq:H_CII13}
\end{align}
{\color{black} For details, see the Appendix.}
   
By going to $k$-space via a Fourier transform and performing an analytic continuation to the entire complex plane, one can now use Eq. (\ref{eq:H_CII13}) to define an $F$-matrix by $F(z) = \sum_n T_n z^n$, in exact analogy to how it was done in Sec. II.B for the AIII symmetry class. We can again use the construction $d(z) = \det F(z)$ for extracting the winding number $\nu = N_z - N_p$, with $d(z)$ well defined also at criticality when the gap is closed (and $d(z)$ has one or several zeros on the unit circle). 

{\color{black} Any first-quantized Hamiltonian} which can be put in symmetry class CII can also be put in AIII (however, with a stronger protection of its topological edge states {\color{black} since now there are many more types of perturbations against which the states are protected, having removed the constraints of time-reversal and particle-hole symmetry)}. One thus expects that the bulk-boundary correspondence that we derived in Sec. II for AIII should still be valid. In other words, one anticipates that the winding number $\nu = N_z - N_p$ obtained for a 1D model in symmetry class CII correctly counts its number of topological edge states, also at criticality. In fact, when discussing the multiband problem for symmetry class BDI in Sec III we relied precisely on this line of argument, grounded in the work by Verresen {\em et al.} \cite{VJP} on two-band models. While the argument is expected to be valid also for CII, the absence of results for two-band models in CII $-$ since none exists in this symmetry class! $-$ may call for a closer look. To this we turn next. However, rather than proving the existence of edge states by an explicit construction for CII, we shall again exploit the bulk-boundary correspondence for critical edge states in AIII. To make the argument formally sound we will make the connection between the CII and AIII winding numbers mathematically manifest, taking off from the four-band Majorana chain in Eq. (\ref{eq:H_CII12}). {\color{black}
This will clarify why the CII winding numbers can take only {\em even} integer values, whereas there is no such restriction for AIII \cite{Chiu2016}. Let us here point out that the generalization to $2n$ bands with $n>2$ becomes cumbersome to handle in the second-quantized Majorana representation. Instead, it is preferable to start directly with a first-quantized single-particle CII Hamiltonian and proceed as for the multiband BDI systems in Sec. III.B, exploiting the results in Sec. II.D for multiband single-particle models in the AIII symmetry class. Again, the calculation is straightforward but long-winded. Since the existence of edge states is fully expected given the analysis of the four-band case, we omit it here.}  

\subsubsection{Connection between CII and AIII winding numbers}
We begin by writing the general CII Hamiltonian in Eq. (\ref{eq:H_CII12}) on first-quantized form, reversing the Nambu procedure from above and by this extracting the corresponding BdG Hamiltonian $H^\text{CII}_\text{BdG}$, call it simply $H$:
 \begin{align} 
\begin{split} 
H =  \sum_{j, n} \big( & t_n (\tau_z \otimes \mathbb{1} - i \tau_x \otimes \sigma_y)   \\
& + \bar{t}_n  (i\tau_z \otimes \sigma_y + \tau_x \otimes \mathbb{1})\big) \otimes |j\rangle \langle j+n| + \text{H.c.}
\end{split}
\label{eq:H_CII15}
\end{align}
We then perform a sequence of unitary transformations acting in the particle-hole and spin spaces, with the transformations given by $U = U_3 U_2 U_1$ with 
$U_1 = \frac{1}{2}(\mathbb{1} + \tau_z)\otimes \mathbb{1} + \frac{i}{2}(\mathbb{1} - \tau_z) \otimes \sigma_y, U_2 = \frac{1}{\sqrt{2}}(\mathbb{1} \otimes \mathbb{1} + i \tau_y \otimes \mathbb{1})$, and $U_3 = \frac{1}{\sqrt{2}}(\mathbb{1} \otimes \mathbb{1} + i \mathbb{1} \otimes  \sigma_x)$. As a result, $H \rightarrow UHU^{-1} = H^\prime$, with  
\begin{align} 
\begin{split} 
H^\prime  =&  \sum_{j, n} \big( t_n (\tau_x \otimes \mathbb{1} + i \tau_y \otimes \mathbb{1})   \\
&+ i\bar{t}_n  (\tau_x \otimes \sigma_z + i \tau_y \otimes \sigma_z)\big) \otimes |j\rangle \langle j + n| + \text{H.c.}
\end{split}
\label{eq:H_CII17}
\end{align}
The chiral symmetry operator $S$ becomes diagonal in the rotated basis and takes the form $S=\sum_j \tau_z \otimes \mathbb{1} |j\rangle \langle j|$ (as is easily verified by checking the identity $SH^\prime S^{-1} = - H^\prime$). Moreover, $H^\prime$ is diagonal in spin space and can be decomposed as
\begin{align} 
\begin{split} 
H^\prime  = \sum_{j, n} \big(& (t_n + i{\color{black}\bar{t}_n})|A, \uparrow, n\rangle \langle B, \uparrow, j + n | \\
& + (t_n - i{\color{black}\bar{t}_n})|A, \downarrow, j\rangle \langle B,  \downarrow, j + n| \big)+ \text{H.c.},
\end{split}
\label{decompH}
\end{align}
where $|A, \sigma, j\rangle$ and $|B, \sigma, j\rangle$ are particle and hole states, respectively with spin $\sigma = \uparrow, \downarrow$, attached to the unit cell with index $j$. 
But this is nothing but two copies, labeled by $\uparrow$ and $\downarrow$, of the general two-band AIII Hamiltonian in Eq. (\ref{eq:H1})! It follows immediately that the winding number, is derivable from $d(z) = \det F(z) = \det \sum_n T^\prime_n z^n$, with
\begin{align} 
\begin{split} 
T_n^\prime = \begin{pmatrix} 
t_n + i\bar{t}_n  &   0 \\
0 & t_n - i\bar{t}_n
\end{pmatrix}.
\end{split}
\label{eq:H_CII18}
\end{align}
We can explicitly rewrite $d(z) = \det F(z) = f^2(z) + g^2(z)$ with $f(z) = \sum_n t_n z^n$ and $g(z) = \sum_n \bar{t}_n z^n$. This result is in perfect agreement with the one obtained above using a Majorana representation, but now we can directly refer to the bulk-edge correspondence proved for class AIII. This establishes that the winding number $\nu = N_z - N_p$ for symmetry class CII in 1D correctly counts the number of topological edge states, also at criticality. {\color{black} As mandated by the tenfold way for the 1D CII symmetry class\cite{Chiu2016}, $\nu$ is restricted to {\em even} integers: Factorizing $d(z) = (f(z) + ig(z))(f(z) - ig(z))$, complex zeros are seen to come in conjugate pairs, with real zeros coming with even multiplicities since $t_n$ and $\bar{t}_n$ are real. (The poles trivially come with even multiplicities.)}     

\subsection{Edge states in four-band gapless \\ CII systems in 1D}

Let us finally address the character of the zero-energy edge states, showing that they are Majoranas. 

Zero-energy states have support on only one sublattice in a basis where the chiral symmetry operator $S$ is represented by a diagonal matrix. This transpired from our analysis in Sec II, and holds quite general \cite{AsbothBook}. Knowing that $S$ is indeed diagonal in the rotated spin-diagonal basis of $H^\prime$, we can therefore consider, without loss of generality, a zero-energy state, call it $|\phi\rangle$, with support on the $A$ sublattice only. When writing Eq. (\ref{decompH}) we associated $A$ with a particle state, $(1\, 0)^T$ [with sublattice $B$ being associated with a hole state, $(0 \,1)^T$]. In this notation, and leaving out the spatial part of $|\phi\rangle$, we thus write $|\phi\rangle = (1\, 0)^T\otimes(a \,b)^T = (a \,b\, 0\, 0)^T$, with $(a\, b)^T$ a spinor in spin space with amplitudes $a$ and $b$. Particle-hole symmetry implies that $C|\phi\rangle$ is also a zero-energy state, where, in the rotated basis, $C=\tau_z \otimes \sigma_x \mathcal{K}$. We can always construct two new zero-energy states by adding and subtracting $\phi$ and $C|\phi\rangle$, in this way obtaining 
$|\phi\rangle_\pm = (u_1^\pm\, u_2^\pm\, 0\, 0)^T$ with $u_1^\pm = a \pm b^\ast$ and $u_2^\pm = b \pm a^\ast$. We now go back to the original basis by carrying out the inverse transformations $|\phi\rangle_\pm \rightarrow |\psi\rangle_\pm = U^{-1}|\phi\rangle_\pm = U_1^\dagger U_2^\dagger U_3^\dagger |\phi\rangle_\pm = \frac{1}{2}(u^\pm_1\!-\!iu^\pm_2\ \ u^\pm_2\!-\!iu^\pm_1 \ \ -u^\pm_2\!-\!iu^\pm_1 \ \ u^\pm_1\!+\!iu^\pm_2)^T$. Exploiting the Nambu spinor representation with $\Psi^\dagger_\text{A} = (c^\dagger_{\uparrow}\, c^\dagger_{\downarrow} \,c_{\uparrow}\, -\! c_{\downarrow})$ (with $\Psi^\dagger_\text{A}$ acting in the composite particle-hole and spin space in any unit cell on the $A$ sublattice), we then express the $|\psi\rangle_\pm$ states in second-quantization as
\begin{align}
\begin{split}
& \frac{1}{2} \Psi^\dagger_A (u^\pm_1\!-\!iu^\pm_2\ u^\pm_2\!-\!iu^\pm_1 \ -\!u^\pm_2 \!+\!iu^\pm_1 \ u^\pm_1\!-\!iu^\pm_2)^T |0\rangle \\
= & \ \frac{1}{2} \big((u^\pm_1\! -\! iu^\pm_2)\,c_{\uparrow}^\dagger \, +\,  (u^\pm_2 \!-\! iu^\pm_1)\,c_{\downarrow}^\dagger \\
- &\,(u^\pm_2\! -\! iu^\pm_1)\,c_{\downarrow} \, -\,  (u^\pm_1 \!-\! iu^\pm_2)\,c_{\uparrow}\big)\, |0\rangle,
 \end{split}
 \label{mode1}
\end{align}
with the composite particle-hole and spin space spanned by $c^\dagger_\uparrow |0\rangle = (1 0 0 0)^T, \,c^\dagger_\downarrow |0\rangle = (0 1 0 0)^T, \,c_\uparrow |0\rangle = (0 0 10)^T$, and $c_\downarrow |0\rangle = (0 0 0 1)^T$. Reading off from Eq. (\ref{mode1}), one immediately identifies the mode operators which correspond to these states:
 \begin{align}
\begin{split}
d_\pm & = (u^\pm_1 - iu^\pm_2)c_{\uparrow}^\dagger + (u^\pm_2 - iu^\pm_1)c_{\downarrow}^\dagger \\
& - (u^\pm_2 - iu^\pm_1)c_{\downarrow} - (u^\pm_1 - iu^\pm_2)c_{\uparrow}. 
 \end{split}
 \label{mode2}
\end{align}
Using that $u^\pm_1\!= \!\pm(u_2^\pm)^\ast$, it follows that $d_\pm\! =\! \pm id_\pm^\dagger$, implying that the zero--energy states thus constructed are Majorana states.    

\section{Robustness of the edge states: a~numerical test}

In the previous sections we showed that the boundary states in unperturbed critical 1D multiband chains are connected to a topological number $\nu$, generalizing the ordinary winding number for gapped systems to critical systems in chiral symmetry classes AIII, BDI, and CII. In gapped systems the edge states are robust to any perturbation from uncorrelated disorder which respects the relevant symmetries and leaves the bulk gap open. Here we numerically verify on a case study that the edge states of  a CII critical chain also exhibit robustness to disorder although the gap is now closed. By a symmetry analysis we identify the chiral symmetry as solely responsible for the protection of the states. A possible mechanism of such protection was proposed in Ref. \onlinecite{Verresen2020} for the case of two-band models, but its generalization to the multiband case has remained somewhat unclear and requires a more thorough investigation.

\begin{figure}[t]
  \centering
   \includegraphics[width=6cm,angle=0]{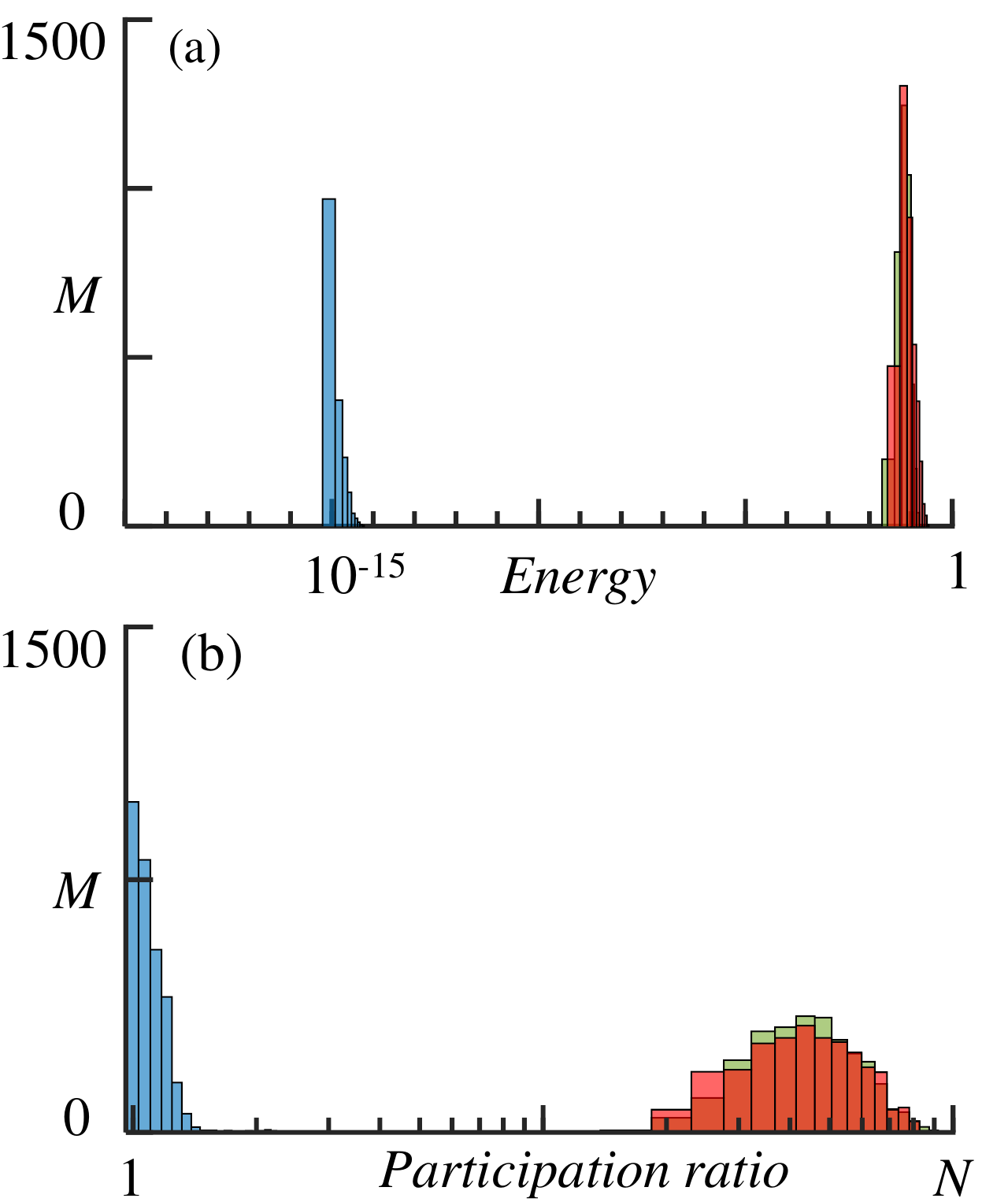}
   \caption{ Number of disorder realizations $M$ (linear scale) corresponding to (a) energy and (b) participation ratio (logarithmic scales) of the four edge states of the gapless CII Hamiltonian in Eq. (\ref{eq:numeric}). The computations cover a total of $10^3$ realizations of each type of the following on-site uncorrelated disorder: proportional to $\tau_x \otimes \mathbf{1}$ (blue, $T$ and $C$ preserving), $\tau_y  \otimes \sigma_y$ (light red, $C$ breaking), and $\tau_y \otimes  \mathbf{1}$ (green, $T$ breaking). (Partial) coloring of bins by dark red corresponds to superposition of light red and green. Each realization of disorder is obtained by summing up on-site perturbations with random amplitudes $\in[-1, 1]$. The number of unit cells is $N = 100$.}
    \label{fig1}
\end{figure}

\begin{figure}[t]
    \centering
    \includegraphics[width=6cm,angle=0]{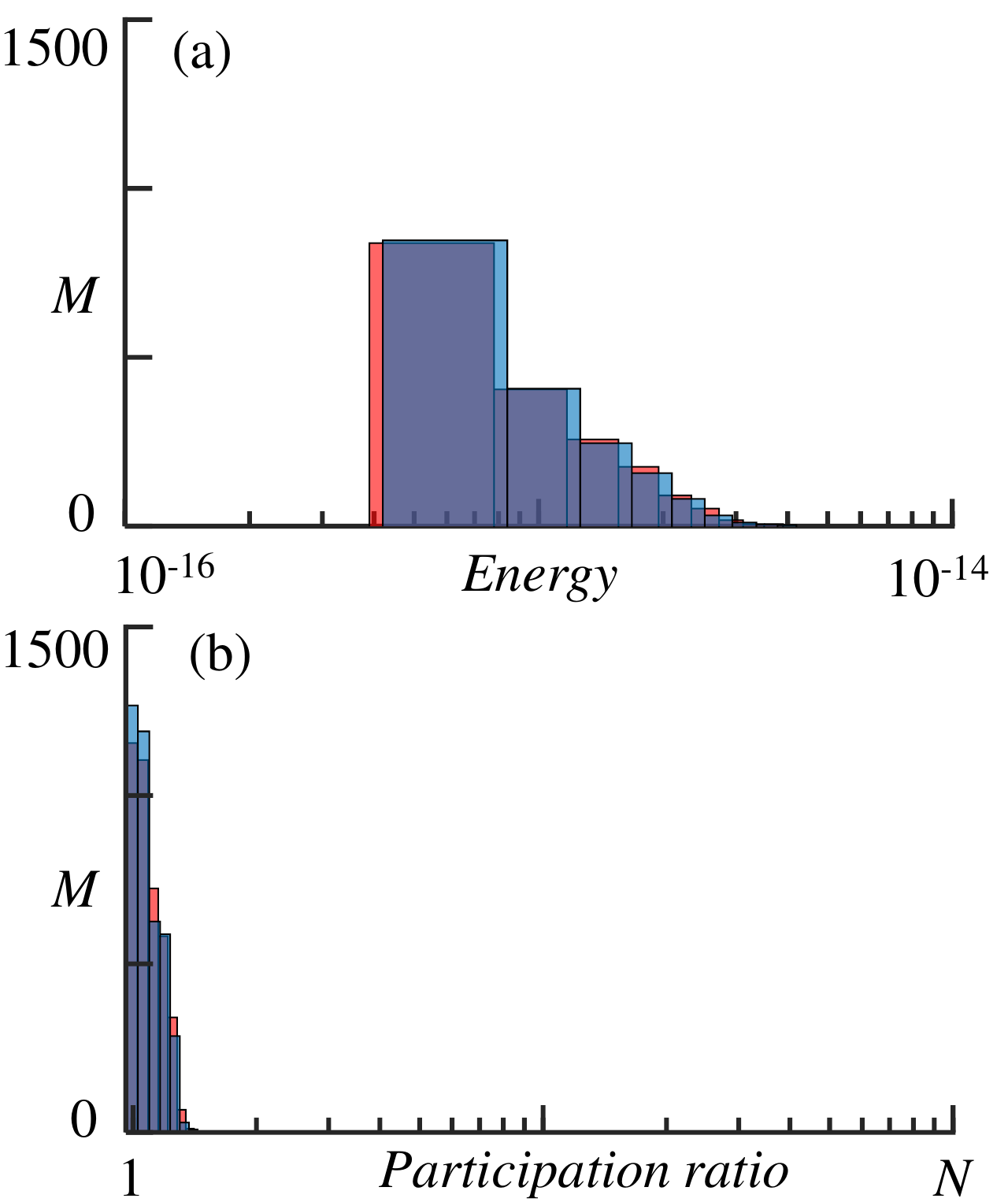}
    \caption{Number of disorder realizations $M$ (linear scale) corresponding to (a) energy and (b) participation ratio (logarithmic scales) of the four edge states of the gapless CII Hamiltonian in Eq. (\ref{eq:numeric}). The computations cover a total of $10^3$ realizations of each type of the following on-site uncorrelated disorder: proportional to $ \tau_x \otimes \sigma_y$ (red) and $\tau_z\otimes \sigma_y$ (light blue), with both perturbations breaking $T$ and $C$ but preserving $S$. (Partial) coloring of bins by dark blue corresponds to superposition of red and light blue. Each realization of disorder is obtained by summing up on-site perturbations with random amplitudes $\in[-1, 1]$. The number of unit cells is $N = 100$.}
     \label{fig2}
\end{figure}

For a numerical test we take a topologically nontrivial critical spinful Majorana chain with open boundaries, ${\cal H}^\text{CII}_{\{1,2\}} = \sum_{n=1}^{2}\! ({\cal H}^\text{CII}_{n}  + \bar{\cal H}^\text{CII}_{n})$, and study how its four edge states behave under disorder respecting the various symmetries of the CII class. {\color{black} (To confirm the criticality of ${\cal H}^\text{CII}_{\{1,2\}}$, insert the nonzero hopping amplitudes of ${\cal H}^\text{CII}_{\{1,2\}}, t_1 \!=\! \bar{t}_1\!=\! t_2\!=\! \bar{t}_2 \!= \!1$, into Eq. (\ref{eq:H_CII18}) and verify that the resulting $d(z)$ has two zeros on the unit circle.)}  As follows from Eq. (\ref{eq:H_CII15}), the Hamiltonian ${\cal H}^\text{CII}_{\{1,2\}}$ can be rewritten in first-quantization as
 \begin{align} 
\begin{split} 
H_{\{1,2\}} =  \sum_{n = \{1, 2\} \atop j=1,2,...,N} \big( & (\tau_z \otimes \mathbb{1} - i \tau_x \otimes \sigma_y)  \\
& + \! (i\tau_z \otimes \sigma_y \!+\! \tau_x \otimes \mathbb{1})\big) \otimes |j\rangle \langle j+n| \!+\! \text{H.c.},  \\
\end{split}
\label{eq:numeric} 
\end{align}
 with $|N+1\rangle$ and $|N+2\rangle$ null states. Having placed $H_{\{1,2\}}$ in symmetry class CII, its relevant symmetries are identified by the time-reversal operator $T =  \sum^N_{j=1} \mathbb{1} \otimes \sigma_y  \,  {\cal K} \otimes |j \rangle \langle j|$, the particle-hole operator $C = \sum^N_{j=1} \tau_y \otimes \mathbb{1} \,  {\cal K} \otimes  |j \rangle \langle j|$, and the chiral symmetry operator $S = \sum^N_{j=1} \tau_y \otimes \sigma_y \otimes  |j \rangle \langle j|$, where ${\cal K}$ is the complex-conjugation operator.

In analogy to gapped symmetry-protected topological systems we are interested to see if the edge states of $H_{\{1,2\}}$ in Eq. (\ref{eq:numeric}) remain localized under symmetry-preserving locally uncorrelated perturbations and, if so, if they stay pinned exactly at zero energy. The localization of the edge states can be quantitatively addressed by calculating the participation ratio (PR) that is defined by $R = 1/\sum_i p_i^{\,2}$, where $p_i$ is the occupation of the Bogoliubov quasiparticle at site $i$, explicitly obtained by summing up all probability amplitudes of the corresponding Nambu spinor. The PR quantifies localization of an eigenstate: A completely localized state has $R = 1$ while a completely delocalized state, such as a plane wave, has  $R = N$ (where $N$ measures the extent of the chain, counting the total number of unit cells). The robustness of an edge state is established if both of the following conditions are satisfied: its energy level remains at zero and the PR is of order unity. 

In Fig.~1 we have perturbed the edge states with random uncorrelated on-site disorder proportional to $\tau_x \otimes \mathbf{1}$ ($T$ and $C$ preserving), $ \tau_y  \otimes \sigma_y$  ($C$ breaking), and $\tau_y \otimes  \mathbf{1}$ ($T$ breaking). The edge states display robustness when both symmetries are preserved (implying that also chiral symmetry, $S = T^{-1}  C$, is preserved) but get destroyed once one of the symmetries, {\color{black} $T$ {\em or} $C$, is broken (implying that also chiral symmetry gets broken \cite{Chiu2016}). Thus, the critical edge states are seen to be well protected in the CII symmetry class. 

It is interesting to inquire what happens if allowing disorder which breaks {\em both} $T$ and $C$ symmetries but leaves chiral symmetry unbroken. This is tantamount to put the Hamiltonian $H_{\{1,2\}}$ in the AIII symmetry class where any perturbation which respects only chiral symmetry is allowed. One may maybe object and say that this assignment of symmetry class is impossible: ``Particle-hole symmetry is a 'built-in'  feature of any second-quantized Hamiltonian expressed in a Nambu spinor basis, implying that the corresponding first-quantized BdG Hamiltonian (like $H_{\{1,2\}}$), as well as any perturbation thereof, is also particle-hole symmetric." However, whereas the use of the Nambu basis does preclude particle-hole symmetry-breaking perturbations in the second-quantized theory, there is no such constraint on perturbations of the BdG Hamiltonian. In other words, once the BdG Hamiltonian matrix has been extracted from the underlying second-quantized theory using a Nambu basis, this matrix defines a single-particle theory which can be put into the AIII symmetry class, allowing for perturbations that break particle-hole symmetry. True, such perturbations do not represent physical symmetry-breaking perturbations of the original unperturbed theory of a mean-field superconductor since they cannot be traced back to a second quantized formulation using the Nambu basis.
However, the maneuver allows us to formally address the question whether critical edge states of a first-quantized Hamiltonian (like $H_{\{1,2\}}$) are robust against perturbations which respect {\em only} chiral symmetry. This is an important issue, independent of whether such perturbations can be realized in an underlying second-quantized theory or not. 

With this as a backdrop,} we now consider perturbations which break both $T$ and $C$ but, different from the cases of Fig.~1, preserve $S$. The results are displayed in Fig.~2 where we have applied random uncorrelated on-site disorder proportional to $\tau_x \otimes \sigma_y$ and $\tau_z\otimes \sigma_y$  ($T$ and $C$ breaking, $S$ preserving), showing that the critical edge states of $H_{\{1,2\}}$ do survive perturbations which respect only chiral symmetry. 

Given these results we conjecture that it is precisely the chiral symmetry $S$ which protects the topological edge states in disordered 1D critical phases.  Numerical examinations of other disorder types and critical chains support this conjecture~\cite{unpublished}. One should here note that while chiral symmetry by definition is indeed the only {\em possible} protecting symmetry for critical phases of AIII, it is {\em a priori} not evident that it actually fulfills this role. However, our numerical results provide strong evidence that it does. 

 \section{Summary}

Building on the work by Verresen {\em et al.} \cite{VJP} on critical two-band BDI models in 1D, we have carried out a study of critical multiband models in any of the 1D chiral symmetry classes AIII, BDI, and CII. We use an approach where the enlarged unit cell responsible for the multiband structure of a model is further extended until one is left with intercell hopping of fermions only between nearest-neighbor auxiliary cells. This allows for a transparent and rigorous analysis of the problem, enabling us to prove the existence of critical edge states for any 1D multiband model belonging to one of the chiral symmetry classes. As in the original work in Ref. \onlinecite{VJP}, the number of such edge states is coded by a topological invariant generalizing the winding number of gapped 1D models, now providing a bulk-boundary correspondence for {\em all} chiral critical phases in 1D. 

A numerical case study of a four-band model in the CII symmetry class $-$ perturbing its Hamiltonian by uncorrelated disorder distributions of different symmetry contents $-$ shows that the robustness of its critical edge states is protected {\em solely} by chiral invariance, with time-reversal and particle-hole symmetries playing no role. We conjecture that this picture is general, with chiral symmetry being the sole protecting symmetry of 1D critical topological edge states not only in the AIII symmetry class (where any perturbation which respects {\em only} chiral symmetry is allowed), but also for any 1D model belonging to the CII or BDI symmetry class. Put differently, we expect that the subsets of the CII and BDI symmetry classes composed of models that support topological critical phases are entirely contained within the AIII symmetry class, demoting $T$ and $C$ to accidental symmetries when it comes to protection of these phases.

Our work may open a path towards a more comprehensive study of symmetry protection of multiband topological phases at criticality, including those of non-translational invariant models in higher dimensions and artificially generated phases from Floquet topological engineering \cite{Harper}. {\color{black} The classification of periodically driven (Floquet) systems at criticality is here a particularly promising direction for exploration: Our \textit{multiband} analysis should be quite useful for treating critical time-periodic systems within Floquet theory -- a theory that is intrinsically multiband due to the repetition of frequency zones~\cite{Shirley, Sambe}. In fact, the Floquet systems represented within frequency domain and truncated at sufficiently large frequency index are mathematically equivalent to time-independent multiband systems~\cite{Shirley, Sambe, Rudner}. Any Floquet chiral system is then anticipated to satisfy the time-independent chiral relation after the truncation, and combined with our results this shows the existence of the topological edge states. Importantly, these arguments are expected to hold also at the closed induced gap (known as the 'anomalous gap') corresponding to the existence of anomalous edge states~\cite{AsbothFL, Rudner}, having no analog in static systems, also at criticality. 

Another important topic to explore is the effect of interactions in multiband critical models. There are already some results \cite{VJP,Verresen2019,Verresen2020,VJPSM} on the robustness of critical topological edge modes against interactions, using density matrix renormalization group (DMRG) and effective field theory methods, but so far only for two bands. The extension to more bands is technically challenging, but we expect that our present work will be of use also here. 

In conclusion, further theoretical work and the growing backdrop of relevant multiband experimental systems \cite{Lutchyn,Stanescu,Setiawan,Samokhin,Mizushima,Mendl,Maslowski} hold up the prospect of some very interesting developments.}
\\ \\ \\ \\

\section*{Acknowledgments}
We thank Eddy Ardonne, Abhishodh Prakash, Christian Sp\aa nsl\"att, and Ruben Verresen for discussions and communications. 
This work was supported by the Swedish Research Council under Grant No. 621-2014-5972 and by the Knut and Alice Wallenberg foundation under grant No. 2017.0157.

\setcounter{equation}{0}
\renewcommand\theequation{A\arabic{equation}}
{\color{black} \section*{\\ APPENDIX: \\CII spinful Majorana chains}}
To obtain all possible CII spinful Majorana chains we first construct all real-space first-quantized Hamiltonian matrices $H$ allowed by the proper time-reversal and particle-hole symmetries and then use a Nambu representation to go to second quantization with complex fermions and, from there, to a representation in spinful Majoranas. 

Recall that the symmetry conditions are $THT^{-1} = H$ and $CHC^{-1} = -H$, with $T^2 = C^2 = -1$. In a real-space representation (with $H$ defined on a chain with $N$ unit cells) we have that 
\begin{equation} \label{symmops}
T = \sum^N_{j=1} \mathbb{1}\! \otimes \!\sigma_y \,\mathcal{K} \otimes |j \rangle \langle j|, \ \ \
C = \sum^N_{j=1} \tau_y \!\otimes \!\mathbb{1}\,\mathcal{K} \otimes |j \rangle \langle j |, 
\end{equation}
with $j$ running over all unit cells, and where $\tau_\alpha$ and $\sigma_\alpha$ are Pauli matrices when $\alpha = x,y,z$ and equal to the $2\times2$ unit matrix $\mathbb{1}$ when $\alpha =0$. The operator ${\cal K}$ effects complex conjugation. For the four-band case, with the real-space Hamiltonian given by a Hermitian $4N\times4N$ matrix, we make the observation that any such matrix can be written as a linear combination over real numbers of matrices 
\begin{equation} \label{Hmatrix}
H_{j, j + n} = (i)^m \tau_\alpha \otimes \sigma_\beta \otimes |j \rangle \langle j \!+ \!n| + \text{H.c.}, \ \ m=0,1,
\end{equation}
with $j$ constrained by $|j+n| \le N$. Examining all 32 such matrices (for any fixed $j$ and $n$), we find that 8 of them respect the $T$ and $C$ symmetries. Writing out only the parts which act in the spin- and particle-hole spaces (suppressing the common $|j\rangle \langle j\!+\!n|$ factors and the Hermitian conjugate to save space), these are:
\\ \\
(a) $\, \,\tau_x \otimes \mathbb{1}$;  \, (b) $\tau_z \otimes \mathbb{1}$;  \ \ (c) $\, \tau_y \otimes \sigma_x ; \, \, \, \, \, \mbox{(d)} \ \ \tau_y \otimes \sigma_z$; \\
(e) $i\mathbb{1}\otimes \sigma_x$; (f) $i\mathbb{1}\otimes \sigma_z$; (g) $i\tau_x \otimes \sigma_y$; (h) $i\tau_z \otimes \sigma_y$.  
\\ \\
\indent To construct the symmetry-respecting second-quantized Hamiltonians ${\cal H}_{j,j+n}$ corresponding to (a) - (h), we introduce four-component real-space Nambu spinors for the $N$ unit cells: $\Psi^\dagger_j = (c^\dagger_{j, \uparrow} \ c^\dagger_{j, \downarrow} \ c_{j, \downarrow}\,  -c_{j, \uparrow})$ and $\Psi_j\!= \!(c_{j, \uparrow} \ c_{j, \downarrow} \ c^\dagger_{j, \downarrow}\,  -c^\dagger_{j, \uparrow})^T$, $j=1,2,...,N$, yielding the $4N$-component Nambu spinors for the full lattice, $\Psi^\dagger = (\Psi^\dagger_1... \Psi^\dagger_j....\Psi^\dagger_{j+n}... \Psi^\dagger_N)$ and $\Psi = (\Psi_1... \Psi_j....\Psi_{j+n}... \Psi_N)^T$ respectively. Inserting the expressions for $H_{j,j+n}$ implied by (a) \!- \!(h), cf. Eq. (\ref{Hmatrix}), into the prescription 
${\cal H}_{j,j+n} = (1/2)\Psi^\dagger H_{j,j+n}\Psi$, we find that only four of them, corresponding to (a), (b), (g), and (h), give a nonzero second-quantized Hamiltonian. In the other four cases, the Hermiticity of $H_{j,j+n}$ when combined with the fermion algebra cancels out the resulting second-quantized expressions, signaling an incompatibility with fermion statistics. 
Listing the four surviving contributions, one finds, from  Eq. (\ref{Hmatrix}) and the table (a)-(h) above, 
\begin{eqnarray} \label{a}
{\cal H}_{j,j+n}^{(a)} & \!=\! & \frac{1}{2} \Psi^\dagger H^{(a)}_{j,j+n} \Psi \nonumber \\
& \!= \!& \frac{1}{2}\Psi^\dagger (\tau_x \otimes \mathbb{1})\otimes(|j \rangle \langle j \!+ \!n| + |j+n \rangle \langle j |) \Psi   \nonumber \\ 
& \!= \!& \frac{1}{2}\Psi^\dagger_{j} (\tau_x \otimes \mathbb{1})\Psi_{j+n} + \frac{1}{2} \Psi^\dagger_{j+n} (\tau_x \otimes \mathbb{1})\Psi_{j} \nonumber \\
& \!= \!& c_{j, \downarrow} c_{j+n, \uparrow} - c_{j, \uparrow} c_{j+n, \downarrow} - c^\dagger_{j, \downarrow} c^\dagger_{j+n, \uparrow} + c^\dagger_{j, \uparrow} c^\dagger_{j+n, \downarrow}, \nonumber \\ 
\end{eqnarray}
and similarly for the other three cases,
\begin{eqnarray} \label{b}
{\cal H}_{j,j+n}^{(b)} & \!= \!& \sum_{\sigma=\uparrow,\downarrow} (c_{j, \sigma}^\dagger c_{j+n, \sigma} + c^\dagger_{j+n, \sigma} c_{j, \sigma}),  \\
\label{g}{\cal H}_{j,j+n}^{(g)} & \!= \!& \sum_{\sigma=\uparrow,\downarrow} (c_{j, \sigma}c_{j+n, \sigma} - c^\dagger_{j, \sigma} c^\dagger_{j+n, \sigma}),   \\
\label{h}{\cal H}_{j,j+n}^{(h)} & \!= \!& c^\dagger_{j, \uparrow}c_{j+n, \downarrow} \!- \!c^\dagger_{j, \downarrow}c_{j+n, \uparrow}\! - \!c^\dagger_{j+n, \uparrow} c_{j, \downarrow} \!+\! c^\dagger_{j+n, \downarrow} c_{j, \uparrow}. \nonumber \\
\end{eqnarray}
In Eqs. (\ref{b}) and (\ref{h}) an immaterial constant (from anticommuting the fermion operators) has been dropped.

Next, we rewrite Eqs. (\ref{a})-(\ref{h}) in terms of spinful Majorana operators, defined by 
\begin{equation} \label{gamma}
\gamma_{j,\sigma} = \frac{1}{2}(c^\dagger_{j,\sigma} + c_{j,\sigma}), \ \ \
\tilde{\gamma}_{j,\sigma} =\frac{i}{2}(c^\dagger_{j,\sigma} - c_{j,\sigma}). 
\end{equation}
\\
By this we obtain
\begin{eqnarray} 
\label{aM} {\cal H}_{j,j+n}^{(a)}\hspace{-0.1cm} & \!=\!& \hspace{-0.17cm}\frac{i}{2}\!(\!\tilde{\gamma}_{j,\uparrow}\gamma_{j+n,\downarrow}\! \!-\! \tilde{\gamma}_{j,\downarrow}\gamma_{j+n,\uparrow}
\!-\! \tilde{\gamma}_{j+n,\downarrow}\gamma_{j,\uparrow} \!+\! \tilde{\gamma}_{j+n,\uparrow}\gamma_{j,\downarrow}\!)  \nonumber \\
\\
\label{bM}{\cal H}_{j,j+n}^{(b)}\hspace{-0.1cm} & \!= \!& \hspace{-0.15cm} \frac{i}{2} \sum_{\sigma=\uparrow,\downarrow} (\tilde{\gamma}_{j,\sigma}\gamma_{j+n,\sigma} + 
\tilde{\gamma}_{j+n,\sigma}\gamma_{j,\sigma}),  \\
\label{gM}{\cal H}_{j,j+n}^{(g)}\hspace{-0.1cm} & \!= \!& \hspace{-0.15cm} \frac{i}{2} \sum_{\sigma=\uparrow,\downarrow} (\tilde{\gamma}_{j,\sigma}\gamma_{j+n,\sigma} - 
\tilde{\gamma}_{j+n,\sigma}\gamma_{j,\sigma}),   \\
\label{hM}{\cal H}_{j,j+n}^{(h)}\hspace{-0.1cm} & \!= \!& \hspace{-0.15cm} \frac{i}{2}(\!\tilde{\gamma}_{j,\uparrow}\gamma_{j+n,\downarrow} \!-\! \tilde{\gamma}_{j,\downarrow}\gamma_{j+n,\uparrow}
\!+\! \tilde{\gamma}_{j+n,\downarrow}\gamma_{j,\uparrow} \!-\tilde{\gamma}_{j+n,\uparrow}\gamma_{j,\downarrow}). \nonumber \\
\end{eqnarray}

By summing Eqs. (\ref{bM}) and (\ref{gM}) and then summing over all unit cells, we obtain a basis $\{ {\cal H}_n^{\text{CII}} \}$ for all four-band {\em spin-preserving} CII Majorana chains: 
\begin{equation} 
{\cal H}^\text{CII}_{n}  =  i \sum_{j, \sigma} \tilde{\gamma}_{j, \sigma} \gamma_{j + n, \sigma}.
\label{eq:SP}
\end{equation}
Analogously, by summing Eqs. (\ref{aM}) and (\ref{hM}) and then again summing over all unit cells, one obtains a basis $\{ \bar{\cal H}_n^{\text{CII}} \}$ for all four-band {\em spin-flipping} CII Majorana chains:   
\begin{equation} 
\bar{\cal H}^\text{CII}_{n}  =  i \sum_{j} (\tilde{\gamma}_{j, \uparrow} \gamma_{j + n, \downarrow} -  \tilde{\gamma}_{j, \downarrow}\gamma_{j + n, \uparrow}).\\
\label{eq:SF}
\end{equation}
It follows that any spinful Majorana chain can be constructed from Eqs. (\ref{eq:SP}) and (\ref{eq:SF}),
\begin{equation} 
{\cal H}^\text{CII}  =  \sum_{n=-\Lambda}^{\Lambda}\! (t_n {\cal H}^\text{CII}_{n}  + \bar{t}_n \bar{\cal H}^\text{CII}_{n}),
\label{eq:SPplusSF}
\end{equation}
with real constants $t_n$ and $\bar{t}_n$, and where $[-\Lambda, \Lambda]$ is the range of couplings between the Majoranas. 

{\color{black} For the purpose of verifying that ${\cal H}^\text{CII}$ has the desired CII symmetries, we introduce the time-reversal and particle-hole symmetry operators ${\cal T}$ and ${\cal C}$, respectively, acting on the Nambu spinors $\Psi_j = \!(c_{j,\uparrow} \ c_{j,\downarrow} \ c^\dagger_{j,\downarrow} \ - c^\dagger_{j,\uparrow})^T$ introduced above and constructed from the unitary parts of the corresponding $T$ and $C$ operators \cite{Chiu2016} in Eq. (\ref{symmops}),
\begin{equation} \label{secondopsT}
{\cal T}\psi_k{\cal T}^{-1} = (U_T)_k^{\ell} \psi_{\ell}, \ \ {\cal T} i {\cal T}^{-1} = -i,
\end{equation}
and
\begin{equation} \label{secondopsC}
{\cal C}\psi_k{\cal C}^{-1} = (U^\ast_C)_k^{\ell} \psi_{\ell}^\dagger.
\end{equation}
Here $U_T = \mathbb{1} \otimes \sigma_y$ and $U_C = \tau_y \otimes \mathbb{1}$ (after having suppressed the extraneous cell index that labels the Nambu spinors), with $\psi_1 = c_{\uparrow}, \psi_2 = c_{\downarrow}, \psi_3 = c^\dagger_{\downarrow},$ and $\psi_4 = - c^\dagger_{\uparrow}$ their common elements in the composite spin and particle-hole space. It follows from Eqs. (\ref{gamma}), (\ref{secondopsT}), and (\ref{secondopsC}) that
\begin{equation} \label{gammaT}
\gamma_{\uparrow} \rightarrow -i\gamma_{\downarrow}, \ \ \gamma_{\downarrow} \rightarrow i\gamma_{\uparrow}, \ \
\tilde{\gamma}_{\uparrow} \rightarrow -i\tilde{\gamma}_{\downarrow}, \ \ \tilde{\gamma}_{\downarrow} \rightarrow i \tilde{\gamma}_{\uparrow}
\end{equation}
under a ${\cal T}$-transformation, while
\begin{equation} \label{gammaC}
\gamma_{\uparrow} \rightarrow i\gamma_{\downarrow}, \ \ \gamma_{\downarrow} \rightarrow -i\gamma_{\uparrow}, \ \ 
\tilde{\gamma}_{\uparrow} \rightarrow -i \tilde{\gamma}_{\downarrow}, \ \ \tilde{\gamma}_{\downarrow} \rightarrow i\tilde{\gamma}_{\uparrow} 
\end{equation}
under a ${\cal C}$-transformation. Reinserting the cell index, one confirms from Eqs. (\ref{eq:SP}), (\ref{eq:SF}), (\ref{gammaT}) and (\ref{gammaC}) that 
\begin{equation} \label{HCII}
{\cal T}{\cal H}^\text{CII}{\cal T}^{-1} = {\cal H}^\text{CII}, \ \ \ {\cal C}{\cal H}^\text{CII}{\cal C}^{-1} = {\cal H}^\text{CII}, \end{equation}
as required.}     
\newpage
 


\begin{thebibliography}{10}
\bibitem{foot}
If $\nu$ is negative, the appearance of edge states at criticality is conditioned by the inequality $\nu < -2c$, with their number per edge given by $|\nu + 2c|^{10}$.
\end{thebibliography}
\end{document}